\documentclass[letterpaper,12pt]{article}
\usepackage{amssymb}

\usepackage{ifpdf}
\ifpdf
	\usepackage[pdftex]{graphicx}
	\usepackage[pdftex,unicode,implicit]{hyperref}

	\hypersetup{
  	pdftitle     = {5D supersymmetric domain wall solution with active hyperscalars and mixed AdS/non-AdS asymptotics}, 
  	pdfkeywords  = {},
  	pdfauthor    = {J. Bellor\'{\i}n and C. Colonnello},
  	pdfcreator   = {pdf\LaTeXe\ with package \flqq hyperref\frqq},
  	pdfproducer  = {pdf\LaTeXe\ with package \flqq hyperref\frqq},
  	pdfpagemode  = UseNone,  
  	pdffitwindow = true,  
  	unicode      = true,
  	plainpages   = true,
  	colorlinks   = true,  
  	citecolor    = blue,  
  	urlcolor     = blue,
  	linkcolor    = blue
	}

	\newcommand{\hepth}[1]{
		arXiv:{\tt\href{http://www.arXiv.org/abs/hep-th/#1}{hep-th/#1}}}

	\newcommand{\arxiv}[1]{
		{\tt\href{http://www.arXiv.org/abs/#1}{arXiv:#1}}}

\else
  \usepackage[dvips]{graphicx}
  \usepackage[unicode,implicit]{hyperref}
  \newcommand{\hepth}[1]{arXiv:{\tt hep-th/#1}}

  \newcommand{\arxiv}[1]{{\tt arXiv:#1}}
  
\fi

\makeatletter
\@addtoreset{equation}{section}
\makeatother

\begin{document}

\begin{flushright}
{\small
Nov $25^{\rm th}$, $2010$}
\end{flushright}

\begin{center}

\vspace{2cm}
{\bf \LARGE 5D supersymmetric domain wall solution with active hyperscalars and mixed AdS/non-AdS asymptotics}
\vspace{2cm}

{\sl\large Jorge Bellor\'{\i}n and Claudia Colonnello}

\vspace{1cm}

{\it Departamento de F\'{\i}sica, Universidad Sim\'on Bol\'{\i}var, Valle de Sartenejas,\\ 
1080-A Caracas, Venezuela.} \\[1ex]
{\tt jorgebellorin@usb.ve, ccolonnello@sinata.fis.usb.ve} \\

\vspace*{2cm}
{\bf Abstract}
\end{center}

\begin{quotation}\small
We find a new supersymmetric 5D solution of $\mathcal{N} = 2$ supergravity coupled to one hypermultiplet that depends only on the fifth dimension (the energy scale in a holographic context). In one asymptotic limit the domain wall approaches to the $AdS_5$ form but in the other one does not. Similarly, the hyperscalars, which are all proportional between them, go asymptotically to a critical point of the potential only in one direction. The quaternionic K\"ahler manifold of the model is the $H^4$ hyperboloid. We use the standard metric of $H^4$ in an explicit conformally flat form with several arbitrary parameters. We argue that the holographic dual of the domain wall is a RG flow of an $D=4$, $\mathcal{N} = 1$ gauge theory acquiring a conformal supersymmetry at the IR limit, which correspond to the $AdS_5$ asymptotic limit.
\end{quotation}

\newpage

\pagestyle{plain}




\section{Introduction}
Supersymmetric solutions of 5D gauged supergravity theories have received a lot of interest by several reasons, in particular by the role they play in understanding the Gravity/Gauge correspondence \cite{varios:adscft}. The foundational scheme in which the correspondence was formulated consists of an $AdS_5 \times S^5$ space-time in the bulk and a $\mathcal{N} = 4$, $U(N)$ Super Yang-Mills Theory in the boundary. In general, it is expected that this seminal Gravity/Gauge correspondence can be extended in order to get the holographic duals of other field theories, perhaps closer to our physical world. Among these extensions, it is interesting to have non-vanishing scalars fields in the bulk, such that the bulk does no longer correspond to the $AdS_5$ vacuum (omitting the compact component) and its holographic dual can be interpreted as a perturbation of the pure Super Yang-Mills Theory. The running of the scalars fields along the fifth dimension (the ``energy'' coordinate) could be regarded as the holographic dual of the RG flow for some coupling constants of the field theory in the boundary. This aspect has been extensively studied in the literature, some of the related papers can be found in Refs.~\cite{varios:holographicflow,Freedman:1999gp,varios:d5dwnohypers,varios:5ddw,varios:stability} and references therein. 

To have the holographic dual of some RG flow it is commonly accepted that the dependence of the bulk scalar fields on the fifth dimension has the form of a kink or domain wall. In an asymptotic direction the scalar fields go to a critical point of the potential, the bulk recovers the AdS form whereas holographically the dual field theory approaches to the Super Yang-Mills theory in the UV regime. In this scheme it is not a mandatory requirement that the bulk approaches to AdS also at the other extreme of the energy coordinate. Similarly, the dual field theory does not necessarily recover the pure Super Yang-Mills theory at the IR. 

Supersymmetric solutions of supergravity are natural candidates for such domain walls because the conditions for preserved supersymmetry can be easily casted as flow equations along the energy coordinate.

Supersymmetric solutions of supergravity with AdS asymptotics are also used in braneworld models \cite{Randall:1999vf}. In these scenarios it is proposed that our observable world is a brane embedded in a higher-dimensional space with large extra dimensions. When the braneworld is formulated as a smooth configuration with a scalar field, then gravity can be trapped around the brane if the high-dimensional space is a domain wall solution interpolating between two AdS extrema. There are several physical requirements that this kind of domain walls must fulfill, in particular the scalar field must be allowed to smoothly run from one critical point of the potential to another with the same energy. There are some no-go theorems that forbid the presence of supersymmetric domain walls of this kind when the supergravity theory is coupled only to vector multiplets \cite{varios:nogo}. For the case of hyperscalars the general known results are less restrictive \cite{Ceresole:2001wi}. Here supersymmetric domain walls suitable for smooth braneworld models have been found using non-homogeneous quaternionic K\"ahler manifolds \cite{Behrndt:2001km}.

Mainly motivated by their relevance in the Gravity/Gauge correspondence, in this work we look for new 5D supersymmetric smooth domain wall solutions with scalar fields. We concentrate ourselves in the $D=5$, $\mathcal{N} = 2$ (eight supercharges) gauged Supergravity coupled to one hypermultiplet. This theory can be useful to find holographic duals of non-maximally supersymmetric field theories at the boundary. The gauged scenario we consider for the supergravity theory is needed in order to find supersymmetric domain wall solutions that have some AdS asymptotics, since there is a potential available for the scalar fields which might play the role of the cosmological constant in asymptotic limits. The supergravity multiplet contains a vector field that allows the gauging of one isometry direction of the quaternionic K\"ahler manifold. This simple matter content is rich enough to provide us with a suitable frame for the analysis of domain wall solutions.

One of the advantages of studying supersymmetric solutions of supergravity is that their characterization has been broadly analyzed. For the $D=5$, $\mathcal{N}=2$ Supergravity and its matter couplings this has been done in Refs.~\cite{varios:classification,Bellorin:2007yp,Bellorin:2008we}. In particular, the characterization of the supersymmetric solutions of the gauged theory coupled to an arbitrary number of hypermultiplets was achieved in Ref.~\cite{Bellorin:2007yp}. Therefore, we may start our analysis by taking the conditions obtained in Ref.~\cite{Bellorin:2007yp} as the set of equations that define the problem of finding supersymmetric solutions. This is a quite fruitful approach, in particular due to the small number of conditions we must solve in order to have a supersymmetry-preserving configuration. Indeed, we shall consider the null case, in which there is roughly one equation for each bosonic field: the metric, vector and the hyperscalars.

To perform explicit computations with hyperscalars one must choose the target manifold. Here the difficulty lies in the involved technology of the quaternionic K\"ahler geometry, in contrast to the very special geometry that arises when the scalars of the vector multiplets are used. With the aim to get a frame in which exact computations can be carried out, the quaternionic K\"ahler manifold we shall use in this work is the 4D hyperboloid $H^4 = SO(4,1)/SO(4)$ (or Euclidean $AdS_4$). Our strategy will consist in casting the standard metric of the hyperboloid as an explicit conformally flat metric by using a coordinate system that depends on six parameters. The parameters are subject to an algebraic constraint due to the fact that the value of the scalar curvature of the manifold is fixed in the gauged supergravity, otherwise they are free. We shall exploit greatly the freedom in the values of the parameters in order to find the solutions. We shall also use the general Killing vector of the hyperboloid as well as its associated momentum map in such coordinates. The general Killing vector has other free parameters associated to each isometry transformation; these can also be adjusted when solving the equations for supersymmetric solutions.

Since we are interested in domain wall solutions, we assume an ansatz in which the space-time metric and the hyperscalars depends only on the fifth coordinate. In addition, we shall assume a conformally flat form for the space-time metric. After we formulate all the equations needed to have a supersymmetric solution, we shall perform the exact integration of the resulting flow-equations obtaining in this way the exact supersymmetric solutions.

This paper is organized as follows: In section 2 we present the action, equations of motion and supersymmetry variations of the $D=5$, $\mathcal{N}=2$ gauged Supergravity coupled to one hypermultiplet, as well as a summary of the general conditions required for supersymmetric solutions that are yielded by the characterization program. In section 3 we define the quaternionic K\"ahler manifold we are going to use. In section 4 we find the supersymmetric solutions by  integrating out the resulting flow-equations.


\section{Supersymmetric solutions of $D = 5$, $\mathcal{N} = 2$ gauged Supergravity coupled to one hypermultiplet}
\label{action}
In this section we present the main definitions of the 5D supergravity as well as the conditions for preserved supersymmetry that we are going to analyze.

The theory for which we are going to look for supersymmetric solutions is the $D = 5$, $\mathcal{N} = 2$ gauged supergravity coupled to one hypermultiplet, with no vector/tensor multiplets. The supergravity multiplet is given by the fields $(e_\mu{}^a , A_\mu , \psi_\mu^i)$ whereas the hypermultiplet is composed of $(q^X , \zeta^A)$. The presence of the vector field $A_\mu$ allows the gauging of an isometry direction of the quaternionic K\"ahler manifold, which is the target manifold of the hyperscalars. The details of this theory can be found in Refs.~\cite{Ceresole:2000jd,Bergshoeff:2004kh}.

The bosonic part of the action is
\begin{equation}
S  =   
\int d^5 x  
\left[ \sqrt{|g|} \left( - R + {\textstyle\frac{1}{2}} g_{XY} \mathfrak{D}_\mu q^{X} \mathfrak{D}^\mu q^{Y}
+ \mathcal{V}(q) 
- {\textstyle\frac{1}{4}} F_{\mu\nu} F^{\mu\nu} \right)
+ {\textstyle\frac{1}{12\sqrt{3}}} \epsilon^{\mu\nu\alpha\beta\gamma}  F_{\mu\nu} F_{\alpha\beta} A_{\gamma}
\right] \,,
\end{equation}
where the scalar potential is defined by
\begin{equation}
\mathcal{V}(q) =
4g^2 ( \vec{P} \cdot \vec{P}
  -{\textstyle\frac{3}{8}} k^{X} k_X ) \,.
\label{potential}
\end{equation}
In this Lagrangian $g_{XY}$ is the quaternionic K\"ahler metric of the non-compact quaternionic K\"ahler manifold. This manifold is also endowed with a quaternionic structure $\vec{J}_X{}^Y$ and an $\mathfrak{su}(2)$ gauge connection $\vec{\omega}_X$. In addition, the general formalism in which the theory is formulated uses a class of vielbein denoted by $f_X{}^{iA}$. The Killing vector $k^X$ determines the isometry of the quaternionic K\"ahler manifold used for the gauging and $\vec{P}$ is its associated momentum map. $\mathfrak{D}_\mu$ is the space-time covariant derivative made with the affine, spin, $A_\mu$ and R-symmetry connections. It acts on the hyperscalars and the gravitino as
\begin{equation}
\begin{array}{rcl}
\mathfrak{D}_\mu q^X & = & \partial_\mu q^X + g A_\mu k^X \,,
\\[1.5ex]
\mathfrak{D}_\mu \mathfrak{D}_\nu q^X & = &
\nabla_\mu \mathfrak{D}_\nu q^X 
+ g A_\mu \partial_Y k^X \mathfrak{D}_\nu q^Y 
+ \Gamma_{YZ}{}^X \mathfrak{D}_\mu q^Y \mathfrak{D}_\nu q^Z \,,
\\[1.5ex]
\mathfrak{D}_\mu \psi^i_\nu & = &
\nabla_\mu \psi^i_\nu 
+ i(\partial_\mu q^X \vec{\omega}_X 
+ {\textstyle\frac{1}{2}} g A_\mu \vec{P}) \cdot \vec{\sigma}_j{}^i \psi^j_\nu \,.
\end{array}
\end{equation}
The corresponding equations of motion for the bosonic fields are 
\begin{eqnarray}
G_{\mu\nu}
- {\textstyle\frac{1}{2}}g_{XY}\left(
\mathfrak{D}_{\mu} q^X \mathfrak{D}_{\nu} q^Y
-{\textstyle\frac{1}{2}} g_{\mu\nu}
\mathfrak{D}_\rho q^X \mathfrak{D}^{\rho} q^Y \right)
+ {\textstyle\frac{1}{2}}g_{\mu\nu}\mathcal{V}
& = & 0\, ,
\label{einsteineom}
\\[1ex]
\mathfrak{D}_\mu \mathfrak{D}^\mu q^X 
- \partial^X \mathcal{V} 
& = & 0 \,,
\label{hypereom}
\\[1ex]
\nabla_\mu F^{\mu\nu}
+ \frac{1}{4\sqrt{3}}  \frac{\epsilon^{\alpha\beta\gamma\mu\nu}}{\sqrt{|g|}} 
F_{\alpha\beta} F_{\gamma\mu} 
+ g k_X \mathfrak{D}^\nu q^X  & = & 0 \,.
\end{eqnarray}
The supersymmetry transformation rules for the fermionic fields, evaluated on
vanishing fermions, are
\begin{equation}
\begin{array}{rcl}
\delta_{\epsilon}\psi^{i}_{\mu} 
& = & 
\mathfrak{D}_{\mu}\epsilon^{i}
-{\textstyle\frac{1}{8\sqrt{3}}} F^{\alpha\beta}
\left(\gamma_{\mu\alpha\beta}-4g_{\mu\alpha}\gamma_\beta\right)
\epsilon^{i}
+{\textstyle\frac{i}{2\sqrt{3}}} g \vec{P}\cdot\vec{\sigma}_{j}{}^{i} \gamma_\mu\epsilon^j \, , 
\\[2ex]
\delta_{\epsilon}\zeta^A 
& = &
{\textstyle\frac{1}{2}}\left( \not\!\!\mathfrak{D} q^X 
+ \sqrt{3} g k^X \right) f_X{}^{iA} \epsilon_i \,.
\end{array}
\label{kse}
\end{equation}

The characterization of the supersymmetric solutions of this theory was done in Ref.~\cite{Bellorin:2007yp}, where it was used the approach of the killing spinor bilinears in ordinary tensorial calculus language. As usual in those studies, one has two families of solutions: the time-like and the null class. Our first step in the task of looking for supersymmetric solutions flowing along the fifth dimension is to determine which of these two families is the most suitable scenario to find such solutions. This can be elucidated by contrasting the conditions imposed directly on the hyperscalars in both cases, Eqs. (3.24) and (3.62) of Ref.~\cite{Bellorin:2007yp}. It turns out that in the  null-case the structure of the differential equation for the hyperscalars is very near to a flow equation (in terms of covariant derivatives), where the flow is mainly governed by the Killing vector $k^X$. On the contrary, a flow-like structure in the analogous equation of the time-like case is not so clear. Due to this, we choose to work with the null-case family.

In the null-case, the supersymmetric solutions are built according to the following recipe (actually, we adopt the somewhat different handling of variables of Ref.~\cite{Bellorin:2008we}): in the space-time one uses two light-cone coordinates, $u$ and $v$, and three transverse coordinates $x^r$. Tensorial objects that have components only in the transverse space are denoted by a hat $\hat{}$. For the space-time metric one chooses two functions $f$ and $H$ and a 3-dimensional transverse metric $\gamma_{{r}{s}}$ together with a Driebein basis $v^{\underline{r}}$ for it, being $\varpi^{\underline{rs}}$ the associated spin connection. An identification between the local symmetry of the Driebein basis and the $SU(2)$ gauge symmetry of the quaternionic K\"ahler manifold is made. For this reason the flat indices $\underline{rs}\ldots$ are carried both by tensors of the transverse space and by objects of the quaternionic K\"ahler manifold. With regard to the other fields, one chooses the components $\hat{A}$ and $A_{u}$ of the vector field $A$ and the four hyperscalars $q^{X}$. All these variables may depend on $u,x^r$ but must be $v$-independent. Moreover, the analysis simplifies greatly if we consider the configurations to be also $u$-independent. This also ensures that the solutions are time-independent. Therefore, from now on we assume this condition, which in turn implies that the direction for the flowing is necessarily contained in the transverse subspace. 

The supersymmetric space-time metric and the 5D vector field are determined in terms of these variables by
\begin{eqnarray}
& & ds^{2} = 
  2 f du (dv + H du + \hat{\omega}) - f^{-2} \gamma_{rs} dx^{r} dx^{s}\,,
\hspace{2em}
\hat{d}\hat{\omega}  = 
  \sqrt{3} f^{-2} \ \hat{\star} \hat{\mathfrak{D}} A_{u} \,, 
\label{eq:rnullmetric}
\\[1ex]
 & & A  =  A_{u} du + \hat{A} \, ,
\hspace{2em}
 F  =  
  \hat\mathfrak{D} A_{u} \wedge du + \hat{F} \,,
\end{eqnarray}
where $\hat{\mathfrak{D}}$ is the transverse covariant derivative made with the affine and spin connections of $\gamma_{rs}$ and the gauge connection $\hat{A}$. The variables chosen must satisfy the following equations for preserved supersymmetry\footnote{The difference on signs between Eqs.~(\ref{eq:rnullquaternionic})-(\ref{eq:rF}) and Eqs.~(3.62)-(3.68) of Ref.~\cite{Bellorin:2007yp} is due to the projection imposed on the Killing spinors. See Ref.~\cite{Bellorin:2008we}}:
\begin{eqnarray}
\varpi^{\underline{rs}} & = & 
2 \varepsilon^{\underline{rst}} (\hat{d}q^X \omega_X^{\underline{t}} +
   {\textstyle\frac{1}{2}} g \hat{A} P^{\underline{t}}  ) 
       - 2\sqrt{3} g f^{-1} P^{[\underline{r}} v^{\underline{s}]}\, ,		
\label{eq:rnullspin}
\\[1ex]
\hat{\mathfrak{D}}_{\underline{r}} q^{X} J^{\underline{r}}{}_{X}{}^{Y} 
& = & 
 - \sqrt{3} g f^{-1} k^{Y}  \, ,
\label{eq:rnullquaternionic}
\\[1ex]
\hat{F}  & = & 
 \sqrt{3} \ \hat{\star} ( \hat{d}f^{-1}
 			- {\textstyle\frac{2}{\sqrt{3}}} g f^{-2} \hat{P} )\,,
 \label{eq:rF}
\end{eqnarray}
where $\hat{P} \equiv P^{\underline r} v^{\underline r}$. We may see that, once we assume the hyperscalars to depend only in one transverse coordinate and discarding the gauge connection, the Eq.~(\ref{eq:rnullquaternionic}) has the structure of a flow equation along this coordinate with an inhomogeneous term proportional to the Killing vector, as we mentioned above. 

A part of the equations of motion of the theory are automatically solved by the supersymmetric configurations. An easy way to see this is to use the Killing Spinor Identities \cite{varios:ksi}. Therefore, the remaining conditions to be imposed on the variables come from the unsolved components of the Maxwell and Einstein equations. These are respectively
\begin{eqnarray}
\hat{\mathfrak{D}}\hat{\star}
(f^{-1} \hat{\mathfrak{D}} A_u)
+ {\textstyle\frac{1}{\sqrt{3}}} \hat{F} \wedge \hat{\mathfrak{D}} A_u 
 + g \hat{P} \wedge \hat{d}\hat{\omega}
	+ {\textstyle\frac{1}{2}} g^2 A_u f^{-3} k^X k_X \mathrm{vol_3} 
 & = & 0 \,, 
\label{eq:rmaxwelleom}
\\[1ex]
\hat{\nabla}^2 H 
+ f^{-1} \hat{\mathfrak{D}}_r A_u \hat{\mathfrak{D}}^r A_u
- {\textstyle\frac{1}{2}} g^2 f^{-3} {A_u}^2 k^X k_X
& = & 0 \,. 
\label{eq:reinsteineom}
\end{eqnarray}
Eqs.~(\ref{eq:rnullspin}) - (\ref{eq:reinsteineom}) define the system we must solve in order to get supersymmetric solutions.


\section{The target for the hyperscalars}
\label{target}
Since we are interested in exact supersymmetric solutions, we must specify a target for the hyperscalars. A quaternionic K\"ahler manifold whose metric is particularly simple is the 4D hyperboloid, which is the coset manifold $SO(4,1)/SO(4)$. We choose this manifold, exploiting the fact that its metric can be put in a conformally flat form. Indeed, it is known that the hyperboloid is the only non-compact 4D quaternionic K\"ahler manifold that admits a conformally flat metric. We shall discuss this result, the discussion will allow us to write the metric in the most general, explicit, conformally flat form, as well as to find the most general form of its Killing vectors. These results will be of great utility when building the explicit solutions, specially due to certain parameters that enter in the general expression of the metric and other geometrical objects of the target and whose values can be adjusted in order to find the solutions.

\subsection{The 4D conformally flat quaternionic K\"ahler manifolds}

Let us discuss this subject as a formal problem. A four-dimensional quaternionic-K\"ahler manifold is a Riemannian manifold restricted by two conditions: Its Weyl tensor is self-dual and its metric is Einstein,
\begin{eqnarray}
 C^{(-)}_{XYZ}{}^V & = & 0 \,,
\\[1ex]
 R_{XY} & = & \kappa g_{XY},
\label{defeinstein}
\end{eqnarray}
where it is assumed $\kappa \neq 0$ as part of the definition. As it is well-known, in supergravity theories the value of $\kappa$ is fixed in terms of the dimension of the manifold. For instance, in $D=5$, $\mathcal{N} = 2$ (Poincar\'e) Supergravity we have
\begin{equation}
\kappa = 
 - {\textstyle\frac{1}{2}} n_H (n_H + 2) \,,
\end{equation}
where $n_H$ is the number of hypermultiplets. This value of $\kappa$ is negative and in particular with one hypermultiplet $\kappa = -3/2$. Since a compact Riemannian manifold with negative scalar curvature has no non-trivial isometries, if we require the existence of at least one isometry the quaternionic K\"ahler manifold must be non-compact.

We are going to prove (locally) that any four-dimensional, conformally flat, quaternionic K\"ahler manifold $\mathcal{M}$ is isometric to either the 4-sphere $S^4$ or the 4-hyperboloid $H^4$ depending on if $\kappa$ is positive or negative respectively. The two cases can be combined into a single metric that depends on six parameters. That is, $\mathcal{M}$ is locally isometric to
\begin{equation}
ds^2_{\mathcal{M}} =
\frac{1}{(a q^X q^X + b_X q^X + c)^2} dq^Y dq^Y \,,
\label{generalmetric}
\end{equation}
where $a$, $b_X$ and $c$ are six constants related to $\kappa$ by the condition
\begin{equation}
b^2 - 4 a c  = - \frac{\kappa}{3}  \,,
\hspace{2em}
b^2 \equiv b_X b_X. 
\label{constraint}
\end{equation}
Notice that in the hypothesis we do not require $\mathcal{M}$ to be homogeneous, hence the result is not a trivial corollary of the standard classification of homogeneous Einstein manifolds.

We start by assuming that the Riemannian metric of $\mathcal{M}$ is conformally flat, such that it can be written in the general conformally flat form,
\begin{equation}
ds^2_{(\mathcal{M})} =
\frac{1}{L^2} dq^X dq^X \,,
\hspace{2em}
L = L(q^1,\ldots,q^4) \,.
\label{conformallyflat}
\end{equation}
Its Weyl tensor vanishes, hence it is automatically self-dual. Therefore, the only remaining condition on $\mathcal{M}$ to be quaternionic-K\"ahler is that it must be Einstein (with $\kappa \neq 0$).

Let us analyze the Einstein condition (\ref{defeinstein}). 
Using the parameterization (\ref{conformallyflat}) this equation becomes
\begin{equation}
2 \partial_{X}\partial_{Y} L 
+ \delta_{XY} \left(
  \partial_{Z}\partial_{Z} L - 3 L^{-1} \partial_Z L \partial_Z L - \kappa L^{-1}
    \right) = 0 \,.
\label{einstein}
\end{equation}
From this equation it is straightforward to deduce that the Hessian of $L$ is diagonal, 
\begin{equation}
\partial_X \partial_Y L = 0 \,,
\hspace{2em}
X \neq Y \,,
\label{offdiagonal}
\end{equation}
and that all its diagonal entries are equal,
\begin{equation}
 \partial_X \partial_X L  =  \partial_Y \partial_Y L
 \hspace{2em} \forall X,Y  \,.
 \label{diagonal}
\end{equation} 
The most general solution to the Eq.~(\ref{offdiagonal}) is $L$ to be an additively-separable function,
\begin{equation}
L = \sum\limits_X Q_X (q^X)  \,,
\label{separable}
\end{equation}
where the index $X$ in $Q_X$ also indicates its argument. By putting this result into Eq.~(\ref{diagonal}) we get the set of equations
\begin{equation}
\frac{d^2 Q_X}{d{q^X}^2} = \frac{d^2 Q_Y}{d{q^Y}^2} 
 \hspace{2em} \forall X,Y  \,.
\label{2derivative}
\end{equation}
Consistency of these equations requires their both sides to be equal to an unique constant. This fact leads us to the general expression of each function $Q_X$: they are second-order polynomials with the same second-order coefficient,
\begin{equation}
Q_X(q^X) = a (q^X)^2 + b_X q^X + c_X
\hspace{2em}
\mbox{(no sum over $X$),}
\end{equation}
where $a$, $b_X$ and $c_X$ are constant. Inserting this result in the expression (\ref{separable}) for $L$ yields
\begin{equation}
L = a q^X q^X + b_X q^X + c 
\hspace{2em}
\mbox{(sum over $X$),}
\label{L}
\end{equation}
where $c \equiv \sum\limits_X c_X$. 

So far we have obtained in (\ref{L}) the form that necessarily $L$ has if the conformally flat manifold is Einstein. The sufficient condition is dictated by Eq.~(\ref{einstein}). By putting the expression (\ref{L}) back into Eq.~(\ref{einstein}) we check that it is solved and the value of $\kappa$ is given by
\begin{equation}
b^2 - 4 a c  = -\frac{\kappa}{3}  \,.
\end{equation}
This proves that any 4D Riemannian, conformally flat, Einstein metric can be written locally as the metric (\ref{generalmetric}), being the constant $\kappa$ given by Eq.~(\ref{constraint}).

We may see clearly the geometric meaning of the metric (\ref{generalmetric}) by using the freedom to perform coordinate transformations. However, it must be taken into account that general coordinate transformations on the target are \emph{not} part of the symmetries of the gauged supergravity theory. In general, different parameterizations of the same metric lead to different physical models. Only transformations corresponding to diffeomorphisms along isometries of the quaternionic K\"ahler metric, which is a smaller group, are allowed as symmetries.

To proceed with the coordinate transformations we need to consider separately the vanishing or not of the parameter $a$: if $a \neq 0$ we may complete squares in the expression of $L$ (\ref{L}),
\begin{equation}
	 L = 
	 a \left(q^X + \frac{b_X}{2a}\right) \left(q^X + \frac{b_X}{2a}\right) 
		+ \frac{\kappa}{12 a} \,,
\end{equation}
where we have used the constraint (\ref{constraint}). In the new coordinates
$p^X = \frac{\sqrt{|\kappa|}}{2\sqrt{3} a} (q^X + \frac{b_X}{2a})$ the metric (\ref{generalmetric}) takes the form
\begin{equation}
	ds^2_{(\mathcal{M})} =
	\frac{12/|\kappa|}{( \mbox{sign}(\kappa) p^X p^X + 1)^2} 
	dp^X dp^X \,.
	\label{stereographic}
\end{equation}
Depending of the sign of $\kappa$, this expression is either the metric of $S^4$ ($\kappa > 0$) or the metric of $H^4$ ($\kappa < 0$), both in stereographic coordinates. If $a = 0$ the conformal factor $L$ is linear in the coordinates $q^X$ and the constraint (\ref{constraint}) fixes the modulus of the ``vector'' $b$, $b^2 = - \kappa/3$. This implies that $\kappa$ cannot be positive and that at least one of the components $b_X$ must be non-vanishing. We can always rotate the coordinate system in order to align one of the axis with $b$. That is, we define the new coordinates	$W = e^b_X q^X$, $V^i = e^i_X q^X$,	where $(e^b , e^i)$ is a constant orthonormal 4D basis, $e^b$ being the unit vector along $b$. Then $L$ becomes linear in $W$ and the inhomogeneous constant can always be absorbed by a shift in $W$. The metric takes the form
\begin{equation}
	 ds^2_{(\mathcal{M})} = 
	 \frac{3 / |\kappa|}{W^2} (dV^i dV^i + dW^2) \,,
	 \label{half}
\end{equation}
which is the metric of $H^4$ in Poincar\'e coordinates. This exhausts all the possibilities for $a$ and also for the sign of $\kappa$. We have seen that if $\kappa > 0$ the metric correspond to $S^4$ with metric given by (\ref{stereographic}) and for $\kappa < 0$ the metric is $H^4$ given either by (\ref{stereographic}) or (\ref{half}).

We shall use the metric (\ref{generalmetric}) of the hyperboloid four our supergravity model. We shall eventually adjust the parameters $a$, $b_X$ and $c$ of this metric in order to be able to find the supersymmetric flow-like solutions. Since $\kappa = -3/2$ for the theory, Eq.~(\ref{constraint}) becomes a constraint on these parameters, which is
\begin{equation}
b^2 - 4 a c  = \frac{1}{2}  \,.
\label{constraintdef}
\end{equation}


\subsection{Geometric properties}
In this subsection we show some geometric properties of the 4D metric (\ref{generalmetric}). In particular, since the isometries of the manifold are highly relevant for the supergravity theory, we find the Killing vectors. We develop the analysis in the general coordinate system in which the metric (\ref{generalmetric}) is written, avoiding to perform any coordinate transformation that could require special values of the parameters.

\subsubsection{Connections and curvature}
The Levi-Civita connection and the Riemann curvature tensor of the metric (\ref{generalmetric}) are
\begin{eqnarray}
 \Gamma_{XY}{}^Z & = &
- 2 L^{-1} \left[ 2 a q^{(X} \delta_{Y)}{}^Z + b_{(X} \delta_{Y)}{}^Z 
   - {\textstyle\frac{1}{2}} \delta_{XY} ( 2 a q^Z + b_Z) \right] \,,
\\[1ex]
R_{XYZ}{}^V & = &
- \frac{2\kappa}{3} g_{Z[X} \delta_{Y]}{}^V  \,.
\end{eqnarray}

Notice that in the Eqs.~(\ref{eq:rnullspin}) - (\ref{eq:reinsteineom}) for supersymmetric solutions the quaternionic structure $\vec{J}_X{}^Y$ and the $\mathfrak{su}(2)$ gauge connection $\vec{\omega}_X$ arise explicitly, in contrast to the original supersymmetry transformations (\ref{kse}), in which the standard vielbein $f_X{}^{iA}$ is needed. This allows us to choose a different vielbein and a convenient quaternionic structure. To the metric (\ref{generalmetric}) we associate the vielbein
\begin{equation}
E_{X}{}^{\underline{X}} = 
 L^{-1} \delta_{X}{}^{\underline{X}}  \,,
\end{equation}
where the underlined indices denote the flat directions. The corresponding spin-connection is given by
\begin{equation}
\Omega_{\underline{RST}} =
- 4 \delta_{\underline{R}[\underline{S}} \delta_{\underline{T}]}{}^{X} 
  \left( a q^{X} + {\textstyle\frac{1}{2}} b_{X} \right).
\label{spinconnectionH4}
\end{equation}
A convenient quaternionic structure in the flat basis is given by the anti-selfdual 't~Hooft symbols,
\begin{equation}
J^r{}_{\underline{X}}{}^{\underline{Y}} = 
\rho^r_{\underline{XY}} =
\left\{ \begin{array}{l}
 \rho^{\underline{r}}_{\underline{st}} = - \epsilon_{\underline{rst}} \,, 
\\[1ex]
 \rho^{\underline{r}}_{\underline{s4}} = \delta_{\underline{rs}} \,,
\end{array} \right.
\end{equation}
whose all components are constant. In the curved basis the components are the same, $\vec{J}_{X}{}^{Y} = \delta_{X}{}^{\underline{X}} \vec{J}_{\underline{X}}{}^{\underline{Y}} \delta_{\underline{Y}}{}^{Y}$.
With this quaternionic structure the anti-selfdual part of the spin-connection (\ref{spinconnectionH4}) can be related to the $\mathfrak{su}(2)$ connection by the general 4D formula $\vec{\omega}_{\underline{X}} = 
\frac{1}{4} \Omega^{(-)}_{\underline{XZ}}{}^{\underline{Y}} \vec{J}_{\underline{Y}}{}^{\underline{Z}}$, which yields
\begin{equation}
\vec{\omega}_{\underline{X}} = 
- \vec{\rho}_{\underline{XY}} 
( a q^{\underline{Y}} + {\textstyle\frac{1}{2}} b_{\underline{Y}} ) \,,
\label{su2connection}
\end{equation}
where for economy we use the symbols $q^{\underline{X}} \equiv q^X$, $b_{\underline{X}} \equiv b_X$.


\subsubsection{Isometries}
We have seen that the 4D conformally flat metric (\ref{generalmetric}) correspond either to $S^4$ if $\kappa > 0$ or $H^4$ if $\kappa < 0$. Assuming a preserved orientation, the isometry groups of these manifolds are $SO(5)$ and $SO(4,1)$ respectively. Therefore, the Killing vectors of the metric (\ref{generalmetric}) corresponds to the action of these groups in the coordinate system in which the metric is written.

Since the metric (\ref{generalmetric}) is conformally flat, an efficient way to obtain its Killing vectors is to start with its \emph{conformal} Killing vectors. These are the same of the Euclidean metric $\delta_{XY}$, which are all well known (in general, conformally equivalent metrics share the same conformal Killing vectors). A conformal Killing vector $k^{X}$ of the metric $g_{XY}$ by definition satisfies
\begin{eqnarray}
 ( \mathcal{L}_k g )_{XY} & = & \Omega_{(g,k)} g_{XY} \,,
\\[1ex]
 \Omega_{(g,k)} & = & 
 {\textstyle\frac{1}{4}} g^{XY} (\mathcal{L}_k g)_{XY} \,.
\label{u}
\end{eqnarray}
Now what we want for the metric (\ref{generalmetric}) are Killing vectors, so we require $\Omega_{(g,k)}$ = 0. This yields the equation
\begin{equation}
  L \partial_{Z} k^{Z} - 4 k^{Z} \partial_{Z} L = 0\,,
\label{condkilling}
\end{equation}
where $L$ is given in (\ref{L}). Therefore, any conformal Killing vector of the Euclidean metric is a Killing vector of the metric (\ref{generalmetric}) iff it satisfies Eq.~(\ref{condkilling}).

The general conformal Killing vector of the Euclidean metric is given by \begin{equation}
 k^{X} =
 \lambda^{Y} ( \delta^{YX} q^Z q^Z - 2 q^{Y} q^{X} )
+ \omega^{XY} q^{Y}
+ \sigma q^{X}
+ \ell^{X}	 \,,
\label{generalkilling}
\end{equation}
where $\omega^{XY}$ is antisymmetric. In each term we recognize the special conformal transformations, rotations, dilatations and translations respectively. This vector correspond to the action of the conformal group $SO(5,1)$.

Now we impose the condition (\ref{condkilling}) on the vector (\ref{generalkilling}). Since this equation must hold at any point, the coefficients of each power of the coordinates $q^X$ must vanish separately. The third-order terms cancel out completely. The second, first and zero-order terms yield respectively the conditions
\begin{eqnarray}
 a \sigma + b_Z \lambda^Z & = & 0 \,,
\label{cond1}
\\[1ex] 
 a \ell^X + {\textstyle\frac{1}{2}} b_Z \omega^ {ZX} + c \lambda^X  & = & 0 \,,
\label{cond2}
\\[1ex]
 b_Z \ell^Z - c \sigma & = & 0 \,.
\label{cond3}
\end{eqnarray}
Therefore, the vector given in (\ref{generalkilling}) is a Killing vector of the metric (\ref{generalmetric}) if the parameters $\lambda^X$, $\omega^{XY}$, $\sigma$ and $\ell^X$ satisfy the conditions (\ref{cond1}) - (\ref{cond3}). These conditions reduce the original conformal group $SO(5,1)$ down to the symmetry groups $SO(5)$ or $SO(4,1)$. We discuss this in Appendix \ref{app:symmetrygroups}. Since Killing vectors are a special case of conformal Killing vectors, we conclude that the above are the general Killing vectors of the metric (\ref{generalmetric}).

The momentum map associated to the general Killing vector (\ref{generalkilling}), which is defined by $\vec{P}=\frac1{2}\vec{J}_X\,^{Y} \nabla_Y k^X$, results
\begin{equation}
 \begin{array}{rcl}
 \vec{P} & = & 
 {\displaystyle\frac{\vec{\rho}_{XZ}}{2L}} \left[ 
 2 \lambda^Y \left( (2 b_V q^Z - q^V b_Z) q^V \delta^{XY} 
 + 4b_Z q^Y q^X \right) 
 + a \omega^{XY} \left( q^V q^V \delta^{YZ} - 4q^Y q^Z \right) \right.
\\[2ex] & &
+ \left. \left( (4c\lambda^X - 2\sigma b_X - 4a\ell^X) \delta^{YZ} + b_Y\omega^{XZ} - 2 b_Z\omega^{XY} \right) q^Y - 2 b_Z \ell^X + c\omega^{XZ}
\right].
\end{array}
\label{generalmomentum}
\end{equation}


\section{The supersymmetric flow-like solutions}
\label{solution}

\subsection{The ansatz for the fields}
To find the solutions we start with switching-off the graviphoton, $\hat{A} = A_u = 0$, which greatly simplifies the Eqs.~(\ref{eq:rnullspin}) - (\ref{eq:reinsteineom}). Indeed, with the vanishing of the graviphoton, which implies the vanishing of the one-form $\hat{\omega}$ given in Eqs.~(\ref{eq:rnullmetric}), the Maxwell equation (\ref{eq:rmaxwelleom}) is automatically solved. The Einstein equation (\ref{eq:reinsteineom}) is solved for any harmonic function $H$ in the transverse space. For simplicity we put $H = 0$.

We identify one of the transverse coordinates, denoted by $w$, as the direction for the flowing and assume that the fields depend only on it. This implies that all the objects $f$, $\gamma_{rs}$, $v^{\underline{r}}$ and $q^X$ depend only on $w$. The transverse coordinates split out as $x^r = (x^i,w)$. With these assumptions the supersymmetric space-time metric given in Eq.~(\ref{eq:rnullmetric}) takes the form
\begin{equation}
ds^{2} = 
  f \left( 2 du dv - f^{-3} \gamma_{rs} dx^{r} dx^{s} \right) \,.
  \label{premetric}
\end{equation}

Our aim is to recover the $AdS_5$ space in some asymptotic direction, which is equivalent to demand that the metric (\ref{premetric}) tends to the $AdS_5$ metric when $w$ goes to some of its ends. By simple inspection we may see that this can be achieved if the function $f$ goes as $\propto 1/w^2$ and at the same time the combination $f^{-3} \gamma_{rs}$ goes as $\delta_{rs}$, such that in that limit the metric (\ref{premetric}) acquires the form
\begin{equation}
ds^{2} \propto 
  \frac{1}{w^2} \left( 2 du dv - dx^{i} dx^{i} - dw^2 \right) \,,
 \label{ads}
\end{equation}
which is the metric of $AdS_5$ in Poincar\'e coordinates. This scheme leads us to make the following ansatz for the transverse-space metric 
\begin{equation}
\gamma_{rs} \, = \, f^3 \delta_{rs}.
\end{equation}
The vielbein associated to this metric are given by
\begin{equation}
v_{r}{}^{\underline r} =
f^{{3}/{2}} \delta_{r}{}^{\underline r} \,,
\end{equation}
where again underlined indices denote flat directions. The transverse-space spin connection $\varpi_{\underline{rst}}$ computed for this veilbein has only the non-vanishing components
\begin{equation}
\varpi_{\underline{ijw}} = 
- {\textstyle\frac{3}{2}} f^{-5/2} f' \delta_{\underline{ij}}.
\label{space-time spin-c}
\end{equation} 
where the prime denotes derivation w.r.t.~$w$. The metric (\ref{ads}) can be written in the coordinate $U = w^{-1}$, which is the coordinate usually associated with the energy scale in Gravity/Gauge correspondence. In general, we assume $0 < w < \infty$.

Our last ansatz for the bosonic fields is to assume that all the hyperscalars are linearly dependent between them,
\begin{equation}
q^{X} = C^{X}Q(w),
\label{unhyperescalar}
\end{equation}
being $Q(w)$ an undetermined function and $C^X$ a constant vector.

After these assumptions we have arrived at the space-time metric
\begin{equation}
 ds^{2} = 
   f(w) ( 2 du dv - dx^{i} dx^{i} - dw^2 ) \,,
   \label{premetric2}
\end{equation}
and the functions $f$ and $Q$ must solve the Eqs.~(\ref{eq:rnullspin}) - (\ref{eq:rF}). By expanding all the spatial components of Eq.~(\ref{eq:rnullspin}) we obtain the system\footnote{The orientation of the space-time has been fixed such that $\epsilon^{ijw}\,=\,\epsilon^{ij}$.}
\begin{eqnarray}
0 & = & Q'\rho^r_{XY} C^{X} b_{Y} \,,
\label{spin1}
\\[1ex]
0 & = & P^{\underline{i}} \,,
\label{picero}
\\[1ex]
f' & = & - {\textstyle\frac{2}{\sqrt{3}}} g f^{3/2} P^{\underline{w}} \,,
\label{f2}
\end{eqnarray}
whereas Eq.~(\ref{eq:rnullquaternionic}) becomes
\begin{equation}	
Q' C^{X} \rho^{\underline{w}}_{XY}  = 
 - \sqrt{3} g f^{1/2} k^{Y}  \,.
\label{Q2}
\end{equation}
By projecting this equation to $C^Y$ we obtain the condition 
\begin{equation}
 C^X k^X = 0 \,.
 \label{ck}
\end{equation}
All the components of Eq.~(\ref{eq:rF}) are implied by Eqs.~(\ref{picero}) and (\ref{f2}).

A way to satisfy Eq.~(\ref{spin1}) for $r = 1,2,3$ without constraining the function $Q$ is to impose that the vector $C$ is proportional to $b$,
\begin{equation}
C^{X} = \beta b_{X},
\label{cb}
\end{equation}
where we assume $\beta, b^2 \neq 0$ in order to get at least one non-trivial hyperscalar.

After condition (\ref{cb}) is imposed, the consistency condition (\ref{ck}) becomes $b_X k^X = 0$. By evaluating this explicitly for the general Killing vector (\ref{generalkilling}), in which we use all our ansatze, we obtain 
\begin{equation}
\beta^2 b^2 \lambda^{Y} b_{Y}  Q^2 
- \beta \sigma b^2 Q 
- \ell^{Y} b_{Y}
= 0 \,.
\end{equation}
Since we want to avoid algebraic conditions on $Q$, we are forced to require the vanishing of the coefficient of each power of $Q$ in this equation, hence we require
\begin{eqnarray}
&& \lambda^{Y} b_{Y}  =  \ell^{Y} b_{Y} = 0 \,, 
\label{cond4}
\\[1ex]
&& \sigma = 0 \,.
\label{cond5}
\end{eqnarray}
We recall that $\sigma$, $\lambda^{X}$, $\ell^{X}$ and $\omega^{XY}$ are not all independent parameters, they must solve Eqs.~(\ref{cond1}) - (\ref{cond3}) in order to define a Killing vector. From these the only one that is not automatically implied by the conditions (\ref{cond4}) - (\ref{cond5}) is the Eq.~(\ref{cond2}). To solve (\ref{cond2}) and (\ref{cond4}) we impose
\begin{equation}
\begin{array}{rcl}
\ell^{X} & = & 
\kappa_\ell \omega^{XY} b_{Y} \,,
\\[1.5ex]
\lambda^{X} & = & 
\kappa_{\lambda} \omega^{XY} b_{Y} \,,
\label{llambda}
\end{array}
\end{equation}
where the constants $\kappa_\ell$ and $\kappa_\lambda$ are subject to	
\begin{equation}
a \kappa_\ell + c \kappa_{\lambda} = \frac{1}{2} \,.
\label{kappas}
\end{equation} 

As a consequence of conditions (\ref{unhyperescalar}), (\ref{cb}), (\ref{cond5}) and (\ref{llambda}) the Killing vector (\ref{generalkilling}) takes the form
\begin{equation}
k^{X}  = 
\omega^{XY} b_{Y} (\kappa_{\lambda} b^2 \beta^2 Q^2  
+ \beta Q + \kappa_\ell ) \,.
\label{prekillingfinal}
\end{equation}
Eq.~(\ref{Q2}) requires the Killing vector to be proportional to $\rho^{\underline{w}}_{XY} b_Y$. In order to match this with the expression (\ref{prekillingfinal}) we impose $\omega^{XY} =  \rho^{\underline{w}}_{XY}$. Notice that we could be more general by imposing a relation of proportionality, but the multiplicative constant in front of $\omega^{XY}$ is just a normalization factor for the Killing vector that can be absorbed by a redefinition of the coupling constant $g$. With all these conditions, the conformal factor (\ref{L}), the Killing vector (\ref{prekillingfinal}) and the momentum map (\ref{generalmomentum}), after a convenient change of parameters, become
\begin{eqnarray}
L & = & 
- \frac{\beta}{4 h^2} \left(
\frac{1}{k} Q^2 - 2 Q + k (1 - h^2)  \right) \,,
\label{LQ}
\\[1ex]
k^X & = & \frac{\beta \rho^{\underline{w}}_{XY} b_Y}{\sqrt{3} g} K \,,
\label{killingfinal}
\\[1ex]
P^{\underline{r}} & = &
- \frac{\sqrt{3}\delta^{\underline{rw}}}{2g} \frac{P}{L} \,,
\label{pfinal}
\end{eqnarray}
where
\begin{eqnarray}
P & \equiv &
\frac{g \beta}{2\sqrt{3} h^2 d} \left(
\frac{1}{k} Q^2 - 2(1-h^2) Q + k (1 - 2 d h^2 - h^2 ) \right) \,,
\label{p1}
\\[1ex]
K & \equiv &
- \frac{\sqrt{3} g} {2d} \left(
\frac{1}{k} Q^2 - 2 d Q - k (1 - 2d - h^2 ) \right) 
\end{eqnarray}
and
\begin{equation}
k = - \frac{1}{2a\beta} \,,
\hspace{2em}
h = \frac{1}{\sqrt{2b^2}} \,,
\hspace{2em}
d = \frac{a}{b^2 \kappa_\lambda} \,.
\end{equation}	
To arrive at these expressions for $P$, $K$ and $L$ the constraint (\ref{constraintdef}) has been solved for $c$ and (\ref{kappas}) for $\kappa_\ell$.

We may see from (\ref{pfinal}) that Eq.~(\ref{picero}) is automatically solved. We are left then with two unsolved equations, (\ref{f2}) and (\ref{Q2}). By substituting the Killing vector (\ref{killingfinal}) and the momentum map (\ref{pfinal}) back in these equations we obtain a system of flow-equations for $f$ and $Q$,
\begin{eqnarray}
f' & = & f^{3/2} \frac{P}{L} \,, 
\label{f}
\\[1ex]
Q' & = & f^{1/2} K \,.
\label{q}
\end{eqnarray}
We remark that $L$, $K$ and $P$ are second-order polynomials of $Q$ and do not depend on $f$.

At this stage, we have supersymmetric flow-like solutions with metric given in (\ref{premetric2}) and hyperscalars given by $q^X = \beta b_X Q$ if we find two functions $f(w)$ and $Q(w)$ that satisfy Eqs.~(\ref{f}) and (\ref{q}). These equations depend on the parameters $\beta$, $k$, $h$ and $d$, which in general are not restricted by any algebraic constraint, apart from $\beta, k, d \neq 0$ and $h > 0$.


\subsection{The roots of the polynomials}
\label{sec:roots}
Before we proceed with solving the flow-equations (\ref{f}) and (\ref{q}), in this subsection we study the zeros of $L$, $K$ and $P$, which are of central importance for these equations. 

The zeros of $L$ represent singularities of the target metric $g_{XY}$. These points must be excluded from the physical domain of the hyperscalars $q^X$ in order to have a finite Lagrangian. Therefore, the flow of $Q$, governed by Eqs.~(\ref{f}) and (\ref{q}), should avoid these points. 

According to Eq.~(\ref{q}), the zeros of $K$ are fixed points for $Q$. This implies that they act as barriers for the flow of $Q$, that is, the value of $Q$ can get arbitrarily close, from above or below, to the zero, but it cannot pass the zero. In this way the flow of $Q$ can be protected against a zero of $L$ if a zero of $K$ is located in a preceding position. Moreover, the flow can be protected against both roots of $L$ by a single root of $K$. We shall see how this happens in the explicit solution in subsections \ref{sec:integration} and \ref{sec:properties}.

From Eq.~(\ref{f}) we may see that if a zero of $P$ is reached in a point $w_0$, then $w_0$ is a critical point of the function $f$. 5D flow-like solutions having such critical points are of special interest in smooth braneworld models, where it is expected that gravity can be trapped around the critical point $w_0$ (if it is a maximum), which correspond to the position of the brane. In order to find such solutions from Eqs.~(\ref{f}) and (\ref{q}), it is convenient to have a zero of $P$ uncovered by the ones of $K$, such that $Q$ effectively passes through that point. Indeed, a typical picture for a smooth braneworld model obtained from Eqs.~(\ref{f}) and (\ref{q}) is a flow of $Q$ starting asymptotically in one zero of $K$, passing through a zero of $P$ (the brane) and ending asymptotically in the other zero of $K$.

We denote the roots of $L$, $K$ and $P$ by $\tilde{Q}_{\pm}$, $\bar{Q}_{\pm}$ and $\hat{Q}_{\pm}$ respectively. They are given by
\begin{equation}
\begin{array}{rcl}
\tilde{Q}_{\pm} & = &
k (1 \pm h ) \,,
\\[1.5ex]
\bar{Q}_{\pm} & = & k ( d \pm \triangle ) \,,
\\[1.5ex]
\hat{Q}_{\pm} & = & k ( 1 - h^2 \pm h \hat{\triangle} ) \,,
\\[1.5ex]
\triangle & \equiv & \sqrt{ ( 1 - d )^2 - h^2 } \,,
\hspace*{2em}
\hat{\triangle} \: \equiv \: \sqrt{ h^2 - 1 + 2 d } \,.
\end{array}
\label{hq}
\end{equation}

Now we study the relative ordering of these roots in function of the space of parameters. It is clear that $k$ plays no role, except by its sign that can invert the relative ordering. Since $h > 0$, the two roots $\tilde{Q}_{\pm}$ of $L$ are real and different between them. In principle, $K$ could have a single real root if $\triangle = 0$, which yields $d = 1 \pm h$, but such a root is equal to $\tilde{Q}_{+}$ or $\tilde{Q}_{-}$. To avoid this, we require the two roots $\bar{Q}_{\pm}$ of $K$ to be real and different between them, that is $\triangle > 0$. This condition leaves us with two sectors in the space of parameters, defined by
\begin{equation}
\begin{array}{l}
\mbox{sector 1:} \hspace{2em} d < 1 - h \,,
\\[1.5ex]
\mbox{sector 2:} \hspace{2em} d > 1 + h \,.
\end{array}
\label{sectors}
\end{equation}
Notice that there is not overlapping between this two sets. A bit of algebra allows us to find that the following orderings hold on each sector:
\begin{equation}
\begin{array}{l}
\mbox{sector 1:} \hspace{2em}
d - \triangle < 1 - h < d + \triangle < 1 + h \,,
\\[1.5ex]
\mbox{sector 2:} \hspace{2em}
1 - h < d - \triangle < 1 + h < d + \triangle \,.
\end{array}
\end{equation} 
This implies that, if $k > 0$, the roots of $K$ and $L$ are ordered as $\bar{Q}_{-} < \tilde{Q}_{-} < \bar{Q}_{+} < \tilde{Q}_{+}$ in sector 1 and $\tilde{Q}_{-} < \bar{Q}_{-} < \tilde{Q}_{+} < \bar{Q}_{+}$ in sector 2. In this case we may see that in sector 1 the flow of $Q$ is protected against the two roots of $L$ if the flow takes values in the range $(-\infty , \bar{Q}_{-})$. That is, given an initial data in which the value of $Q$ belongs to this range, then the whole flow driven by Eqs.~(\ref{f}) and (\ref{q}) never reaches $\tilde{Q}_{\pm}$. Similarly, in sector 2 the flow is protected in the range $( \bar{Q}_{+} , +\infty )$. If $k < 0$ the relative ordering of the roots is completely reversed in each sector, such that the flow is protected in sector 1 in the range $( \bar{Q}_{-} , +\infty )$ and in the range $( -\infty , \bar{Q}_{+} )$ for the sector 2. In all cases the flow also avoids the remaining root of $K$. This fact, which is a consequence of having always one root of $L$ between the two roots of $K$, places an inconvenient to find smooth braneworld solutions.

From this analysis we conclude that the values of $h$ and $d$ must fall into one the sectors defined in (\ref{sectors}) and that in each sector, for any sign of $k$, there is always a range available for the flow of $Q$ protected against the roots of $L$. 

By simple inspection we may see that the zeros of $P$ may or may not belong to these ranges depending on the values of $h$ and $d$.

Finally, we comment that the roots $\bar{Q}_{\pm}$ of $K$ are critical points of the scalar potential $\mathcal{V}$ given in Eq.~(\ref{potential}), hence the value of the hyperscalars at these zeros correspond to vacuum solutions. This was to be expected since these roots are also the zeros of the Killing vector, which in turn are critical points for the scalar potential. This last relationship can be shown to hold on general grounds by using the technology of the quaternionic K\"ahler geometry on the derivatives of Eq.~(\ref{potential})\footnote{The converse is not true in general: there can be critical points of the potential that are not zeros of the Killing vector. To see this we simplify the potential by imposing the condition $q^X = C^X Q$, which is equivalent to saying that there is only one hyperscalar in the functional space. Then the potential (\ref{potential}) becomes
\[
 \mathcal{V} = 
 \frac{1}{4L^2} \left( 12 P^2 - \frac{\beta^2}{h^2} K^2 \right) \,,
\]
which can be brought to the form
\[
 \mathcal{V} = 
 6 W^2 - \frac{18 h^2}{\beta^2} L^2 \left(\frac{ dW }{ dQ } \right)^2 \,,
\]
where $ W = P/ \sqrt{2} L $ is the superpotential \cite{Freedman:1999gp,Ceresole:2001wi,varios:superpotencial}. This potential in general has four different critical points; only two of them correspond to the zeros $\bar{Q}_{\pm}$ of the Killing vector, which are also the critical points of the superpotential $W$.}.


\subsection{The exact solutions}
\label{sec:integration}

Let us start the analysis of the system (\ref{f}) - (\ref{q}) by solving the Eq.~(\ref{q}) for $f$,
\begin{equation}
f^{1/2} = \frac{Q'}{K} \,,
\label{solf}
\end{equation}
and then putting this expression into Eq.~(\ref{f}). This yields a second-order differential equation for $Q$, which is
\begin{equation}
Q'' + \frac{P^{(3)}}{P^{(4)}} (Q')^2 = 0 \,,
\label{q2}
\end{equation}
where $P^{(4)}$ and $P^{(3)}$ are fourth- and third-order polynomials of $Q$ respectively,
\begin{equation}
\begin{array}{rcl}
P^{(3)}(Q) & \equiv & {\displaystyle 
- L {\displaystyle\frac{dK}{dQ}} 
- \frac{P}{2}  \,, }
\\[1.5ex]
P^{(4)}(Q) & \equiv & L K \,.
\end{array}
\label{p3p4}
\end{equation}
Note that $P^{(4)}$ has four different real roots: the ones of $L$ and $K$.

Eq.~(\ref{q2}) can be completely integrated. To achieve this, we first note that the two terms of this equation can be grouped in a total derivative after multiplying it by the integration factor $\exp\left(\int dQ \frac{P^{(3)}}{P^{(4)}} \right)$, such that Eq.~(\ref{q2}) becomes
\begin{equation}
\left[ \exp\left(\int dQ \frac{P^{(3)}}{P^{(4)}} \right) Q' \right]' = 0 \,.
\end{equation}
This equation can be brought to the form
\begin{equation}
\left[ \int dQ \exp\left(\int dQ \frac{P^{(3)}}{P^{(4)}} \right) \right]'' = 0 
\end{equation}
and now it can be integrated twice w.r.t.~$w$, yielding
\begin{equation}
\int dQ \exp\left(\int dQ \frac{P^{(3)}}{P^{(4)}} \right) = \ell_1 w + \ell_2 \,,
\label{q2integrated}
\end{equation}
where $\ell_1$ and $\ell_2$ are two arbitrary integration constants. 

Since $P^{(4)}$ has four different real roots the integral of the integration factor can be easily obtained: if $\mathcal{Q}_\alpha$ denotes the four roots of $P^{(4)}$, then
\begin{equation}
\int dQ \frac{P^{(3)}}{P^{(4)}}  = 
 \ln\left[ \prod\limits_\alpha ( Q - \mathcal{Q}_\alpha )^{r_\alpha} \right]\,,
\label{ln}
\end{equation}
where
\begin{equation}
r_\alpha \equiv 
\left[ P^{(3)} \left( \frac{d P^{(4)}}{d Q} \right)^{-1} \right]_{\mathcal{Q}_\alpha} \,.
\label{defr}
\end{equation}
By putting this result back into Eq.~(\ref{q2integrated}) we finally arrive at the equation
\begin{equation}
\int dQ \prod\limits_\alpha ( Q - \mathcal{Q}_\alpha )^{r_\alpha} = 
\ell_1 w + \ell_2 \,.
\label{integral}
\end{equation}

Once the integral of the l.h.s.~of Eq.~(\ref{integral}) has been performed, this equation becomes an algebraic equation for $Q$ and $w$. By solving for $Q$ in terms of $w$ one obtains the solution of the differential equation (\ref{q2}). Finally, the solution $f(w)$ is obtained by substituting $Q(w)$ into Eq.~(\ref{solf}).

Let us go further in the analysis of the integral in Eq.~(\ref{integral}). Actually, we do not know the general solution of this integral for arbitrary values of the exponents $r_\alpha$. Alternatively, we can look for solutions for some specific values of $r_\alpha$.

We start with evaluating the general expressions of the exponents $r_\alpha$ in terms of the parameters. We recall that the four roots $\mathcal{Q}_\alpha$ are $\bar{Q}_{\pm}$ and $\tilde{Q}_{\pm}$. We denote by $\bar{r}_{\pm}$ the two of the exponents $r_\alpha$ that correspond to $\bar{Q}_{\pm}$ and by $\tilde{r}_{\pm}$ the ones of $\tilde{Q}_{\pm}$. From Eqs.~(\ref{defr}) and (\ref{p3p4}) and using the explicit expressions of the second-order coefficients of $P$, $K$ and $L$ we obtain
\begin{eqnarray}
\bar{r}_{\pm} & = &
- 1 \mp \frac{ 2k (\bar{Q}_{\pm} - \hat{Q}_+) (\bar{Q}_{\pm} - \hat{Q}_-) }
  { 3 ( \bar{Q}_{\pm} - \tilde{Q}_+ ) ( \bar{Q}_{\pm} - \tilde{Q}_- )
             ( \bar{Q}_+ - \bar{Q}_- )} \,, 
\\[1ex]
\tilde{r}_{\pm} & = &
\mp \frac{ 2k (\tilde{Q}_{\pm} - \hat{Q}_+) (\tilde{Q}_{\pm} - \hat{Q}_-) }
{ 3 ( \tilde{Q}_{\pm} - \bar{Q}_+ ) ( \tilde{Q}_{\pm} - \bar{Q}_- ) 
      ( \tilde{Q}_+ - \tilde{Q}_- )} \,.
\end{eqnarray}
After substituting the values of all the roots given in (\ref{hq}) these exponents take the form
\begin{eqnarray}
\bar{r}_{\pm} & = & 
- 1 \mp \frac{1}{3\triangle} \left( 1 + \frac{ h^2 }{ d - 1 \pm \triangle } \right) \,,
\label{barr}  
\\[1ex]
\tilde{r}_+ & = & \tilde{r}_- \: = \:- \frac{1}{3}.
\label{tilder}
\end{eqnarray}
Curiously, the value of $\tilde{r}_{\pm}$ is fixed and, in addition, from (\ref{barr}) it is straightforward to obtain the identity
\begin{equation}
 \bar{r}_{+} + \bar{r}_{-} = - \frac{4}{3} \,.
\label{sumbarr}	
\end{equation}

By using the value of $\tilde{r}_{\pm}$ and the identity (\ref{sumbarr}) solved for $\bar{r}_{-}$ in Eq.~(\ref{integral}), and performing the change of variable given by 
\begin{equation}
 u = (Q - \bar{Q}_-)^{-1} \,,
 \label{uq}
\end{equation}
we obtain
\begin{equation}\label{uintegral}
- \int du \frac{ ( 1 - (\bar{Q}_+ - \bar{Q}_-) u )^{\bar{r}_+} }
         {\left[ ( 1 - (\tilde{Q}_+ - \bar{Q}_-) u )
                 ( 1 - (\tilde{Q}_- - \bar{Q}_-) u ) \right]^{1/3}} 
                 = 
                 \ell_1 w + \ell_2  \,.
\end{equation}
The integrand of this equation can be simplified thanks to the fact that the linear polynomial arising in the numerator is proportional to the derivative of the 2nd-order polynomial of the denominator. Indeed, if we define the polynomials of $u$
\begin{equation}
\begin{array}{rcl}
p_1 & \equiv & 1 - (\bar{Q}_+ - \bar{Q}_-) u  \,,
\\[1.5ex]
p_2 & \equiv &
 ( 1 - (\tilde{Q}_+ - \bar{Q}_-) u ) 
 ( 1 - (\tilde{Q}_- - \bar{Q}_-) u ) \,,
\end{array}
\end{equation}
and use (\ref{hq}), then we obtain that they satisfy the following identity
\begin{equation}
 \frac{d p_2}{du} = 2 k ( d - 1 - \triangle) p_1 \,.
\end{equation}
Therefore, Eq.~(\ref{uintegral}) can be written as
\begin{equation}
\frac{- 1}{[ 2 k (d - 1 - \triangle )]^{ \bar{r}_{+} }}
   \int \frac{du}{ {p_2}^{1/3}} \left( \frac{d p_2 }{du} \right)^{\bar{r}_{+}} 
= \ell_1 w + \ell_2 \,.
\label{pintegral}
\end{equation}

Clearly, for the special case of $\bar{r}_{+} = 1$ the integrand in (\ref{pintegral}) becomes a total derivative, such that the integration can be performed automatically. On this basis we impose the condition $\bar{r}_{+} = 1$ and perform the integration, obtaining
\begin{equation}
{p_2}^{2/3} = \ell_1 w + \ell_2 \,,
\label{p2}
\end{equation} 
where any global multiplicative constant has been absorbed in $\ell_1$ and $\ell_2$.

We parenthetically analyze the consequences of having fixed the value of the exponent $\bar{r}_+$, whose expression in terms of $h$ and $d$ is given in (\ref{barr}). We regard the condition $\bar{r}_{+} = 1$ as an algebraic equation for $d$ that depends on $h$. Its solution is\footnote{In despite of its appearance, the function $\bar{r}_+ (d)$ with fixed $h$ is injective; thus the condition $\bar{r}_+(d) = 1$ has only one solution for $d$.}
\begin{equation}
 d = 
 \frac{5}{24} \left( 5 - \sqrt{ 1 + 24 h^2 } \right) \,.
 \label{d}
\end{equation}
Now we must determine whether this value for $d$ falls in one of the allowed sectors defined in (\ref{sectors}). We find that the condition for the sector 1 is satisfied by (\ref{d}) whenever $h$ is different from one. Therefore, condition (\ref{d}) and $h \neq 1$ ensure that if the flow of $Q$ starts in the protected range for the sector 1 then it always takes values in that range for all $w$.

We may solve the algebraic Eq.~(\ref{p2}) for $u$ in terms of $w$ and then return to the original variable $Q(w)$ by reversing the change of variables (\ref{uq}). Since $p_2$ is a second-order polynomial of $u$, we find two different solutions, which we denote by $Q_{\pm} (w)$, given by  
\begin{equation}
Q_{\pm}(w) =
 \frac{ 2 k \triangle} 
 { 1 \pm \left( ( \ell_1 w + \ell_2 )^{3/2} + m \right)^{1/2} }    
 + \bar{Q}_{-} \,,
 \label{solutionq}
\end{equation}
where
\begin{equation}
m \equiv \frac{1 - d - \triangle}{1 - d + \triangle}.
\label{m}
\end{equation}
To arrive at these expressions we have again redefined $\ell_1$ and $\ell_2$ in order to absorb any constant multiplying the combination $\ell_1 w + \ell_2$. 

It can be seen that $0 < m < 1$ for the sector 1 of the space of parameters. Thus, to ensure the reality of $Q_{\pm}(w)$ we impose $\ell_1 > 0$ and $\ell_2 \geq 0$. Moreover, with these conditions the function $Q_+(w)$ is free from singularities in all the domain of $w$. For $Q_-(w)$ there is still the possibility of a singularity in the point $\hat{w}$ that satisfies 
\begin{equation}
 \hat{w} = \frac{ ( 1 - m )^{2/3} - \ell_2 } { \ell_1 } \,.
\end{equation}
This point is automatically excluded from the domain of $w$ if $\ell_2$ satisfies $ \ell_2 > ( 1 - m )^{2/3} $, hence we impose this condition when dealing with $Q_-(w)$.

From the form of the expression (\ref{solutionq}) it is easy to determine whether the flows of the functions $Q_{\pm}(w)$ fall in some of the protected ranges. We recall that the protected ranges for the sector 1 are $(-\infty , \bar{Q}_- )$ if $k > 0$ and $(\bar{Q}_- , +\infty)$ if $k < 0$. From (\ref{solutionq}) we may see that if $k > 0$ then $Q_+(w) > \bar{Q}_-$ and $Q_-(w) < \bar{Q}_-$ for all $w$. Therefore, $Q_-(w)$ is the only solution that takes values in the protected range for $k > 0$. For $k < 0$ we obtain that $Q_+(w) < \bar{Q}_-$ and $Q_-(w) > \bar{Q}_-$, hence it is again $Q_-(w)$ the function that takes values in the protected range.

We obtain the function $f(w)$ by plugging (\ref{solutionq}) into Eq.~(\ref{solf}). This yields the same result both for $Q_+(w)$ and $Q_-(w)$, which is
\begin{equation}
 f(w) =
  \frac{3 d^2 {\ell_1}^2 }{ 16 g^2 \triangle^2 }
   \frac{ \ell_1 w + \ell_2 }
   { \left( ( \ell_1 w + \ell_2 )^{3/2} + m \right)^{2} } \,.
\end{equation}
This function is real, positive definite and finite in all the domain of $w$ provided that $\ell_1$ and $\ell_2$ satisfy the same conditions imposed for the functions $Q_{\pm}(w)$.

In summary, we have found a new supersymmetric flow-like solution for the gauged $D=5$, $\mathcal{N} = 2$ Supergravity coupled to one hypermultiplet with the hyperscalars taking values in a physically allowed set. The quaternionic K\"ahler metric is the metric of the $H^4$ hyperboloid parameterized as
\begin{equation}
 ds^2_{\mathcal{M}}  =  {\displaystyle \frac{1}{L^2} dq^X dq^X \,,}
\hspace{2em}
 L = a r^2 + b_X q^X + c \,,
\end{equation}
where $c = (4a)^{-1} (b^2 -1/2)$, $b^2 = b_X b_X $, and the quaternionic structure is $\vec{J}_X{}^Y = \vec{\rho}_{XY}$. The Killing vector we use for the gauging together with its momentum map are obtained from Eqs.~(\ref{generalkilling}) and (\ref{generalmomentum}) by restricting the parameters arising in these expressions according to
\begin{equation}
 \begin{array}{lll}
  \lambda^X = \kappa_\lambda \rho^{\underline{w}}_{XY} b_Y \,,
  \hspace{2em} &
  \ell^X = \kappa_\ell \rho^{\underline{w}}_{XY} b_Y \,,
  \hspace{2em} &
  {\displaystyle \kappa_\ell = \frac{1}{2a} (1 - 2 c \kappa_\lambda) }\,,
\\[1.5ex]
  \omega^{XY} = \rho^{\underline{w}}_{XY} \,,
  &
  \sigma = 0 \,. &
 \end{array}
\end{equation}
The solution is given by
\begin{equation}
\begin{array}{l}
ds^2 = 
 f(w) (\eta_{\hat{\mu}\hat{\nu}} dx^{\hat{\mu}} dx^{\hat{\nu}} - dw^2 ) \,,
\\[1.5ex]
q^X = b_X Q(w) \,,
\hspace{2em}
 A = 0 \,,
\\[1.5ex]
f(w) =
  {\displaystyle \frac{3 d^2 {\ell_1}^2 }{ 16 g^2 \triangle^2 }
   \frac{ \ell_1 w + \ell_2 }
   { \left( ( \ell_1 w + \ell_2 )^{3/2} + m \right)^{2} }} \,,
\\[2.5ex]
Q(w) =
 {\displaystyle \frac{ \triangle / a} 
 { \left( ( \ell_1 w + \ell_2 )^{3/2} + m \right)^{1/2} - 1 }
 + \bar{Q}_{-} \,,
\hspace{2em}
\bar{Q}_- = -\frac{1}{2a} ( d - \triangle ) } \,,
\\[2.5ex]
m = {\displaystyle \frac{1 - d - \triangle}{1 - d + \triangle} } \,,
\hspace{2em}
\triangle = \sqrt{(1-d)^2 - h^2} \,,
\\[2ex]
d = {\displaystyle \frac{5}{24} ( 5 - \sqrt{ 1 + 24h^2 } ) \,,
\hspace{2em}
h = \frac{1}{\sqrt{2b^2}} \,. }
\end{array}
\label{wholesolution}
\end{equation}
There is also the restriction $\kappa_\lambda = a/b^2 d$. This solution has two integration constants, $\ell_1$ and $\ell_2$, which are subject to the conditions $\ell_1 > 0$ and $\ell_2 > (1 - m)^{2/3}$ in order to ensure the reality and finiteness of the solution for $0 < w < \infty$. It also depends explicitly on the target parameters $a$ and $b_X$, which are required to satisfy $b^2 \neq 1/2$. 

\subsection{Properties of the solution}
\label{sec:properties}

\subsubsection{Asymptotics and the shape of the domain wall}
We now analyze the asymptotic behavior of the solution. To this end we prefer to work with the coordinate $y$ related to $w$ by $ w = e^{\lambda y} / \lambda $, $\lambda > 0$. The space-time metric and the function $Q$ take the forms
\begin{equation}
\begin{array}{l}
ds^2 = 
 f(y) (\eta_{\hat{\mu}\hat{\nu}} dx^{\hat{\mu}} dx^{\hat{\nu}} 
 - e^{2\lambda y} dy^2 ) \,,
\\[1.5ex]
f(y) =
  {\displaystyle \frac{3 d^2 {\ell_1}^2 \lambda^2}{ 16 g^2 \triangle^2 }
   \frac{ \ell_1 e^{\lambda y} + \ell_2 }
   { \left( ( \ell_1 e^{\lambda y} + \ell_2 )^{3/2} + m \right)^{2} } } \,,
\\[2.5ex]
Q(y) =
 {\displaystyle \frac{ \triangle / a } 
 { \left( ( \ell_1 e^{\lambda y} + \ell_2 )^{3/2} + m \right)^{1/2} -1 }
 + \bar{Q}_{-} }\,.
\end{array}
\label{solutiony}
\end{equation}
Here we have redefined $\ell_1$ with respect to (\ref{wholesolution}) by $\ell_1 \rightarrow \ell_1 \lambda$.

As $y$ approaches to $+\infty$, the functions $f(y)$ and $Q(y)$ behave as
\begin{equation}
 \begin{array}{rcl}
  f(y) & = & 
  {\displaystyle \frac{ 3 d^2 \lambda^2 }{16 g^2 \triangle^2}
  e^{-2\lambda y} 
  + \mathcal{O}(e^{-3\lambda y}) } \,,
\\[2.5ex]
  Q(y) & = &
  {\displaystyle \bar{Q}_-
  + \frac{\triangle}{a \ell_1^{3/4}} e^{-3\lambda y / 4}
  + \mathcal{O}(e^{-3\lambda y / 2}) } \,.
\end{array}
\end{equation}
The function $Q(y)$ goes asymptotically to the root $\bar{Q}_-$, which is a critical point of the scalar potential, and the space-time metric approaches the $AdS_5$ form,
\begin{equation}
 ds^2 =
 \frac{ 6 \lambda^2 }{ | \Lambda |}
 ( e^{-2\lambda y} \eta_{\hat{\mu}\hat{\nu}} dx^{\hat{\mu}} dx^{\hat{\nu}} 
  - dy^2 ) \,,
\hspace{2em}
 \Lambda = - \frac{32 g^2 \triangle^2}{d^2} \,.
\end{equation}

\begin{figure}[t]
\begin{center}
\includegraphics[scale=0.30]{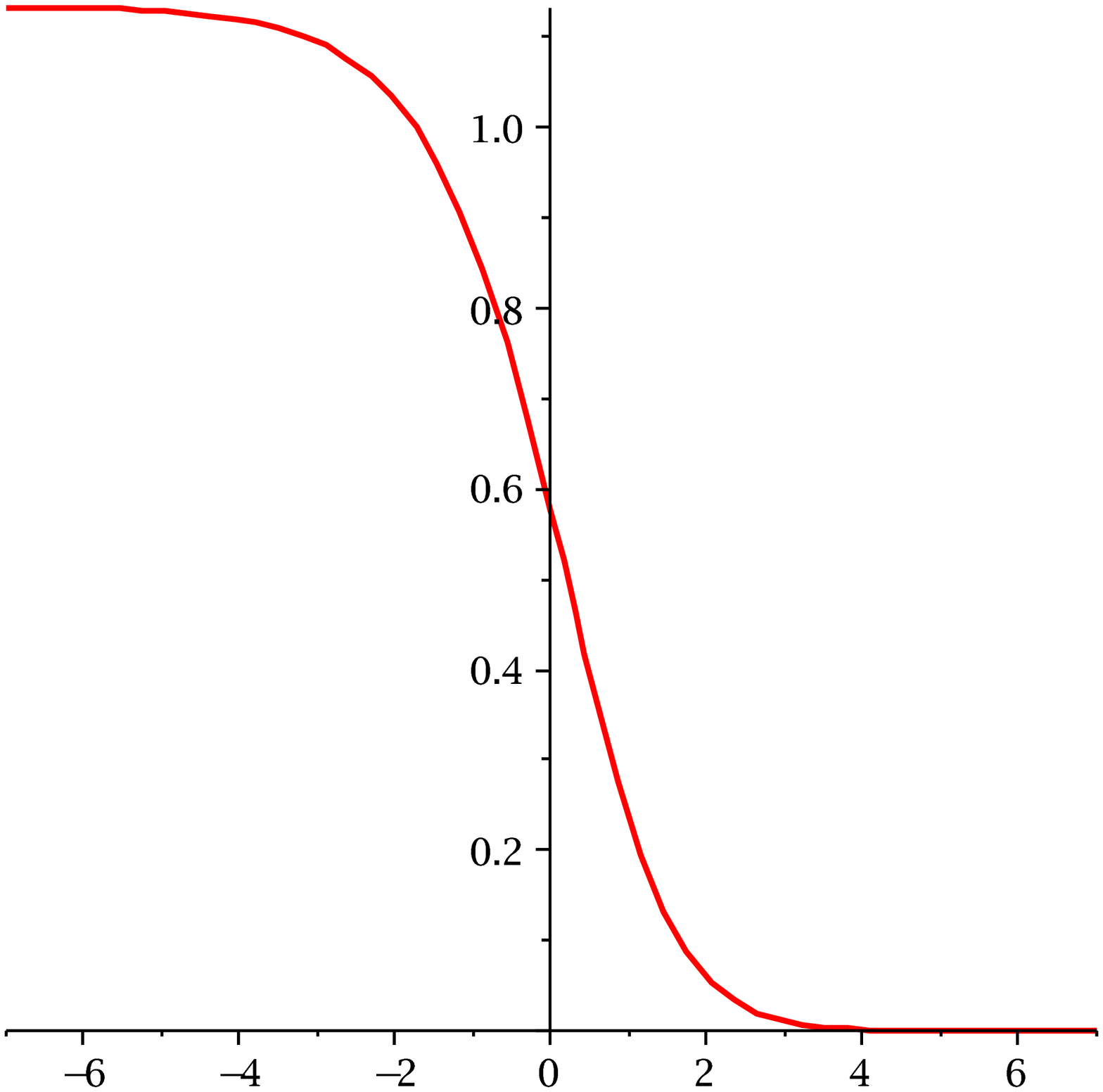}
\hspace{4em}
\includegraphics[scale=0.30]{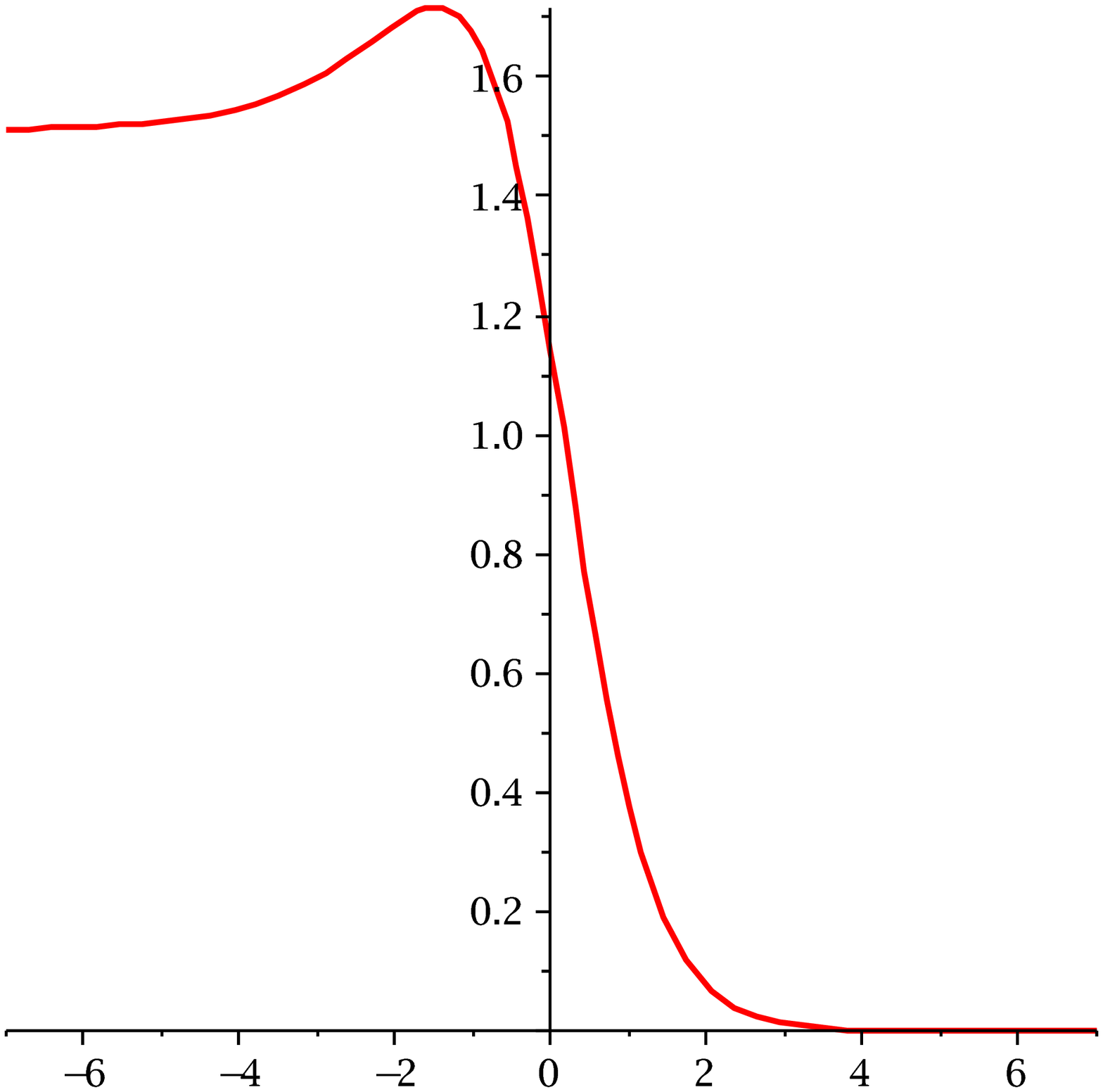}
\caption{(a) A monotonically decreasing function $f(y)$. (b) The function $f(y)$ with a maximum. In both figures the parameters $\beta$, $\lambda$, $\ell_1$ and $g$ have been set to unity, $a=-1$ and $h = 2$. For plot (a) we set $\ell_2 = 1.32$ and for plot (b) $\ell_2 = 0.32$.}
\end{center}
\label{figuraf}
\end{figure}

\begin{figure}[t]
\begin{center}
\includegraphics[scale=0.30]{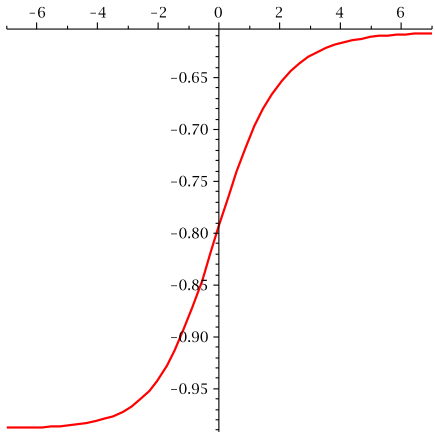}
\end{center}
\caption{The function $Q(y)$. The parameters $k$, $\beta$, $\lambda$ and $g$ have been set to unity, $a = -1$,  $h=2$ and $\ell_2 = 1.32$.}
\label{fig:q}
\end{figure}

Now when $y \rightarrow -\infty$ we obtain
\begin{equation}
 \begin{array}{rcl}
  f(y) & = & 
  {\displaystyle \frac{ 3 d^2 \ell_1^2 \lambda^2 }
  {16 g^2 \triangle^2 \ell_2^{1/2} \delta } \left[
  1 + \frac{ \ell_1 (m - \ell_2^{3/2}/2 ) }{\ell_2 \delta}
  e^{\lambda y} \right]
  + \mathcal{O}(e^{2\lambda y}) } \,,
\\[2.5ex]
  Q(y) & = &
  {\displaystyle 
  Q_{-\infty}
  + \frac{ 3 \triangle \ell_1 \ell_2^{1/2} }
  { 4 a \delta ( \delta^{1/2} - 1 )^2 }
  e^{\lambda y}
  + \mathcal{O}(e^{2\lambda y }) } \,,
\\[2.5ex]
  Q_{-\infty} & \equiv & 
  {\displaystyle \bar{Q}_- + \frac{\triangle}{a ( \delta^{1/2} - 1 ) } }\,,
\end{array}
\label{minusinfinity}
\end{equation}
where $\delta = \ell_2^{3/2} + m$. We may see that in this limit the function $Q(y)$ tends again to a constant value. However, in this case the constant $Q_{-\infty}$ does not correspond to a critical point of the potential. Consequently, the space-time metric does not approach the $AdS_5$ form. At first sight, this asymptotic behavior could be surprising, since one normally expects that if the hyperscalars go asymptotically to constant values, these values must correspond to critical points of the scalar potential and the metric should approach the $AdS_5$ metric. Actually, a careful analysis on the Einstein and hyperscalar equations (\ref{einsteineom}) and (\ref{hypereom}) reveals that this is not the case for the limit $ y \rightarrow -\infty $ due to an effect of the coordinate system. The first term of Eq.~(\ref{hypereom}) contains inverse-metric factors (the $g^{yy}$ factor) that diverge when $ y \rightarrow -\infty $. As a consequence, although $Q(y)$ goes to a constant, this term does not vanish, it goes instead to a finite value and cancels out completely the second term of Eq.~(\ref{hypereom}), the non-zero derivative of the potential. Similarly, all the three terms of the $\hat{\mu}\hat{\mu}$-components of Eq.~(\ref{einsteineom}), including the energy-momentum term of the hyperscalars, go to finite, non-zero values and cancel out between them. In the $yy$-component the three terms go simultaneously to zero in the form $\sim e^{2\lambda y}$. The mentioned divergences are mere effects of the coordinate system. Indeed, by substituting the expansion for $f$ given in Eqs.~(\ref{minusinfinity}) into the line element (\ref{solutiony}) an making a simple redefinition of the fifth coordinate we may see that the leading-order metric correspond to the 5D Minkowski space-time.

As we mentioned in section \ref{sec:roots}, the polynomial $P$ may or may not have a zero in the protected range for the flow of $Q(y)$. This implies that, depending on the values of the parameters, there are two cases for the function $f(y)$: monotonically decreasing or with a maximum. To have a maximum, however, it is not sufficient for the zero of $P$ to be located in the protected range where the flow of $Q(y)$ falls, it must be located between $Q_{-\infty}$ and $\bar{Q}_-$. In particular, this effect can be easily controlled by adjusting the value of $\ell_2$, which increases/decreases the absolute value of $Q_{-\infty}$ while leaves $\bar{Q}_-$ unaffected. In Fig.~1 we show the plots of the function $f(y)$ for the cases when it has a maximum and when it does not. In Fig.~2 we show the increasing version of the domain wall corresponding to the function $Q(y)$.

There are several parameters that can be used to control the shape of the domain wall, both for the metric and the hyperscalars. For example, $\ell_2$ can also be used to reduce the separation between $\bar{Q}_-$ and $Q_{-\infty}$, thus reducing the slope of the wall (and also excluding the maximum for $f$). This feature could be important for the holographic interpretation of the solution since the supergravity approximation in the bulk side is trustable only for low curvatures.


\subsubsection{On the holographic dual of the solution}
Finally, as we mentioned in the Introduction, it would be interesting to understand the holographic dualization of our regular solution \ref{wholesolution}, specifically by relating it to a RG flow of some 4D gauge field theory. Of course, to arrive at a definitive conclusion about what the exact dual gauge theory is, one must compute quantum correlation functions. However, we can gain some insight on the character of the dual gauge theory by looking into the symmetries of the supergravity solution, that is, the preserved supersymmetries and the isometries, in particular at the asymptotic limit at which the AdS vacuum solution is recovered. We devote this section to discuss this.

As a preliminary, we point out that in the bulk we are dealing with a supersymmetric solution. Normally, one expects the perturbative stability of such a solution since it can be related to a BPS state of the underlying quantum theory. This point of view is the usual one in supersymmetric domain walls solutions, even in non-supersymmetric ones, see, for example, Refs.~\cite{varios:stability}.

We firstly need to make contact with the usual parameterization of the domain walls/RG flows used in the Gravity/Gauge correspondence. The fifth coordinate $U$ associated to the energy scale is related to the coordinate $y$ we are using by $U \propto e^{-\lambda y}$. Thus, the IR regime of the dual gauge theory correspond to the $y = +\infty$ limit and the UV regime to the $y = -\infty$ limit. As regards to the coordinate $w$ in which the solution was found in previous sections, we have $U \propto w^{-1}$. Therefore, $w$ may be regarded as an inverse running for the energy scale.

We have seen that at the limit $y = +\infty$ the metric of the solution approaches to the $AdS_5$ solution whereas the scalar fields become constant. This $AdS_5$ is one of the maximally supersymmetric solutions of the theory, since the hyperscalars go to one of the extrema of the scalar potential $\mathcal{V}$ for which the Killing vector vanishes, $k^X = 0$. Hence we have that at the $y = +\infty$ limit the solution preserves all of the eight supersymmetries of the theory and has an isometry group that can be interpreted as a conformal group of transformations acting on the 4D boundary of the $AdS_5$ space. We are led to think that this limit is dual to an $D=4$, $\mathcal{N} = 1$ gauge theory acquiring a conformal supersymmetry at the IR, hence doubling the number of supersymmetries.

The $y = -\infty$ limit, which correspond to the UV regime of the dual gauge theory, must be analyzed carefully due to the subtleties of the coordinate system. Actually, it is better to return to the coordinate $w$ since with it the 5D metric (\ref{wholesolution}), evaluated at the limit $w = 0$, gets the standard form of the Minkowski space-time in Cartesian coordinates. One may check that at this limit the derivative of the hyperscalar $Q$ goes to a constant, non-zero value,
\begin{equation}
 \partial_w Q = 
 \frac{ 3 \triangle \ell_1 \ell_2^{1/2} }
  { 4 a \delta ( \delta^{1/2} - 1 )^2 } \,.
\end{equation}
This constant is exactly the coefficient of the first correction to $Q$ in expansion (\ref{minusinfinity}). It is worth mentioning that this 5D Minkowski solution must not be confused with a vacuum solution of the theory since we have nonzero contributions to the energy-momentum tensor of the hyperscalars at each order in $w$, including the zero-order, as well as nonzero contributions to the scalar potential. Indeed, in this model the critical points of the scalar potential correspond only to solutions with non-zero cosmological constant.

By evaluating the Killing spinor equations (\ref{kse}) on the solution we deduce that they admit a Killing spinor that satisfies the usual projection for 5D supersymmetric domain walls of the $\mathcal{N}=2$ theory,
\begin{equation}
 i \sigma^{\underline{w}}{}_i{}^j \epsilon^j 
 = \gamma^{\underline{w}} \epsilon^i \,.
\end{equation}
For generic null-class supersymmetric solutions there are three of such projections (only two independent), one for each direction of the transverse space \cite{Bellorin:2007yp}. However, in the case of domain walls only one projection is needed due to the dependence in only the fifth coordinate\footnote{In general, there is one additional projection given in terms of a light-cone gamma matrix. This is neither needed for the domain wall.}. This implies that the whole solution preserves one half of the supersymmetries; that is, four real supercharges. This amount of preserved supersymmetry remains unaltered at the UV limit but it is doubled at the IR.

On the basis of these considerations we believe that the 4D gauge theory dual to the whole domain wall solution is an $\mathcal{N} = 1$ gauge theory without conformal supersymmetry in general. Because of the fact that both the enhancement of the supersymmetry and the arising of the conformal structure hold at the IR rather than the UV, it seems that the complete, dual gauge theory can be obtained by making an \emph{explicit} symmetry breaking in some $\mathcal{N} = 1$ conformal supersymmetric theory. The breaking must be done in such a way that the symmetry-breaking terms can be neglected at the IR. This mechanism would be similar to the Lifshitz mechanism of breaking explicitly the Lorentz invariance by adding noncovariant high-order terms to a covariant field theory, such that the Lorentz symmetry is recovered at an IR fixed point. This mechanism was originated in condensed matter theory and recently has been adapted to a gravitational action in order to obtain a renormalizable theory of quantum gravity \cite{Horava:2009uw}. If the correspondence we are proposing works, the $\beta$-function of the symmetry-breaking operator can be related to the $w$-derivative of the hyperscalar $Q$.

As an application, we may use the method of Ref.~\cite{Freedman:1999gp} to obtain a c-function associated to the RG flow for the conformal anomaly of the conformal gauge theory at the boundary. The coefficient of the conformal anomaly is given by the value of the c-function at the IR limit. The c-function of Ref.~\cite{Freedman:1999gp} is defined by
\begin{equation}
 \mathcal{C} = \frac{\mathcal{C}_0}{(A')^3} \,,
\label{cfunction}
\end{equation}
where $A(z)$ is the function determining the domain wall when it is casted in the form
\begin{equation}
 ds^2 = 
 e^{2A(r)} \eta_{\hat{\mu}\hat{\nu}} dx^{\hat{\mu}} dx^{\hat{\nu}} - dr^2 \,,
\label{standarddw}
\end{equation}
and $\mathcal{C}_0$ is a constant. This constant may be evaluated at the IR limit by following the approach of Ref.~\cite{Henningson:1998gx}. Of course, here we cannot give to $\mathcal{C}$ the interpretation given in Ref.~\cite{Freedman:1999gp} as a function for domain walls interpolating between the coefficients of the conformal anomaly at several fixed points because we have the AdS geometry only at the IR limit.

After the coordinate change needed to bring the metric (\ref{wholesolution}) to the form (\ref{standarddw}), 
\begin{equation}
 r = - \sigma^{-1} \ln ( (\ell_1 w + \ell_2)^{3/2} + m ) \,,
 \hspace{2em} \sigma \equiv - 2\sqrt{3} g \triangle / d \,,
\label{coordz}
\end{equation}
($d$ is always negative, hence $\sigma$ is positive) we find that the metric of the solution is given in terms of the function
\begin{equation}
 A(r) = \frac{1}{3} \ln \left( e^{-\sigma r} - m\right) + \sigma r \,.
\end{equation}
The first and second derivatives of this function are
\begin{equation}
\begin{array}{rcl}
  A' & = & 
  {\displaystyle \frac{\sigma}{3} \left( \frac{2 - 3 m e^{\sigma r}}
  {1 - m e^{\sigma r}} \right) } \,,
  \\[2ex]
  A'' & = & {\displaystyle \frac{ - m \sigma^2 e^{\sigma r}}
  {3(1 - m e^{\sigma r})^2} } \,.
\end{array}
\label{A''}
\end{equation}
These functions are free from singularities in the whole domain of $r$, which is bounded from above by a finite value of $r$, as can be deduced from (\ref{coordz}). We may assume that the function $A$ has no critical points in the domain of $r$. This is equivalent to require that the function $f$ of the solution (\ref{wholesolution}) has no critical points, which can always be imposed by restricting the space of parameters of the solution, as we have indicated in the previous section. Therefore, we have a monotonic domain wall. Under this condition, it can be checked from Eqs.~(\ref{A''}) that $A' > 0$ and $A'' < 0$, thus the function $A'$ is always decreasing in $r$. Since $\mathcal{C}$ is related to $A'$ by (\ref{cfunction}) and the IR limit correspond to $r \rightarrow -\infty$, we obtain that the $\mathcal{C}$-function is monotonically decreasing as $r$ runs towards the IR (assuming a positive $\mathcal{C}_0$). In Ref.~\cite{Freedman:1999gp} it was shown that this behaviour is a general property of domain walls that satisfy a weaker energy condition, which in turn is satisfied by domain walls that preserve the 4D Poincar\'e invariance, as in our case.


\section{Conclusions}
We find new exact supersymmetric solutions of the $D=5$, $\mathcal{N}=2$ gauged supergravity coupled to one hypermultiplet. We use the 4D hyperboloid as the target quaternionic K\"ahler manifold for the hyperscalars. We based our analysis in solving the conditions known from the characterization program for the supersymmetric solution of the theory \cite{Bellorin:2007yp} under a flow-like ansatz. One of the solutions we have found is completely regular and the hyperscalars take values in a physically allowed set. This solution is a domain wall that depends only on one noncompact spatial coordinate of the space-time, the fifth coordinate. The asymptotics of the domain wall is mixed: at one extreme of the fifth coordinate it approaches to a maximally supersymmetric $AdS_5$ vacuum solution, whereas at the other direction the hyperscalars tends to a nontrivial configuration and the metric acquires a Minkowski form. This Minkowski plus active hyperscalars configuration is a solution of theory, but not a vacuum solution and it preserves one half of the supersymmetries. The solution as a whole preserves one half of the supersymmetries, as usual in domain wall solutions. We have used a coordinate system in the quaternionic K\"ahler manifold in which the standard metric of the hyperboloid acquires an explicit conformally flat form, but leaving various free parameters. In the space of these parameters, including the ones of the Killing vector used for the gauging, we could find a sector in which the regular solution is located and free from singularities.

It would be interesting to understand completely the holographic dualization of our regular domain wall solution. Actually, to answer this question, quantum computations in the context of the Gravity/Gauge correspondence must be performed. However, upon its asymptotic behavior, we have proposed the supersymmetry/conformal structure content of the 4D gauge theory dual to our solution. The enhancement of the full supersymmetries at the AdS extreme, which correspond to the IR regime of the dual theory, leads us to believe that the holographic dual correspond to a RG flow of a $D=4$, $\mathcal{N} = 1$ gauge theory acquiring a conformal supersymmetry at the IR, but not at the UV. Thus, the gauge theory doubles the number of supersymmetries at the IR. Since the enhancement of the symmetries happens at the IR, one may guess that the dual gauge theory can be obtained from an explicit symmetry-breaking of some $\mathcal{N}=1$ superconformal theory, instead of a spontaneous breaking. The symmetry-breaking terms should be negligible at the IR. It would be interesting to determine which operator can be used to deform the superconformal theory in order to get the dual theory of our solution. The $\beta$-function corresponding to such an operator can be related to the dependence of the hyperscalars of our solution in the fifth dimension. As a feasible application of our solution, we have determined the form of the $\mathcal{C}$ function following the method of Ref.~\cite{Freedman:1999gp}, obtaining that it is a monotonically decreasing function towards the IR.

Our regular domain wall has some similarities with others solutions found in the literature, in particular with those analyses devoted to the five dimensional supergravities. In the five dimensional theories coupled to scalar fields and focused in the Gravity/Gauge correspondence there have been found several supersymmetric domain walls, some concrete examples can be found in Refs.~\cite{Freedman:1999gp, varios:d5dwnohypers, varios:5ddw}. Our work is particularly close to the analyses in Refs.~\cite{varios:5ddw}, where it was studied the same $D=5$, $\mathcal{N}=2$ gauged supergravity coupled to hyperscalars and in some cases also with active scalars of the vector multiplets. These studies, however, were focused in curved domain walls; that is, when the solution does not preserve the 4D Poincar\'e invariance. Another difference is that in the domain walls analyzed in these references both asymptotic limits correspond to AdS vacuum solutions (although not necessarily maximal supersymmetry at both limits), hence the dual theory acquires conformal supersymmetry both at the UV and the IR. This behaviour coincides with the scenario studied in Ref.~\cite{Freedman:1999gp}, where it was used the $\mathcal{N} = 8$ theory to analyze AdS-AdS domain walls. Those authors provided a complete holographic interpretation in terms of a gauge field theory flowing from a conformal point to another due to a deformation, as in the Leigh-Strassler mechanism \cite{Leigh:1995ep}. As we mentioned, our solution  seems to be a different scenario since we have AdS asymptotics only in at the IR limit.


\section*{Acknowledgments}
The authors wish to thank Nigel Hitchin for his invaluable help on the topic of non-compact quaternionic K\"ahler manifolds.


\appendix

\section{Conventions}
We take the action of the theory from Ref.~\cite{Bergshoeff:2004kh}, setting $\kappa = 1/\sqrt{2}$. For the space-time we use the signature $(+----)$. Both in the space-time and target spaces we use underlined indices to denote tangent flat directions. The connection and its curvature tensors are
\begin{eqnarray}
 \Gamma_{\alpha\beta}{}^\gamma & = &
 \frac{1}{2} g^{\gamma\rho} ( 2 \partial_{(\alpha} g_{\beta)\rho}
 - \partial_\rho g_{\alpha\beta} )\,,
\\[1ex]
 R_{\alpha\beta\mu}{}^\nu & = & 
 2 \partial_{[\alpha} \Gamma_{\beta] \mu}{}^\nu
 - 2 \Gamma_{[\alpha| \mu}{}^\rho \Gamma_{|\beta] \rho}{}^\nu \,,
\\[1ex]
 R_{\mu\nu} & = & R_{\alpha \mu \nu }{}^\alpha \,.
\end{eqnarray}
The AdS space-time has a negative cosmological constant. When the scalar potential of the Lagrangian becomes constant, it is related to the cosmological constant by $\mathcal{V} = - 2 \Lambda$.

In the target space the conventions are similar. The Levi-Civita connection we use and its curvature tensors are
\begin{eqnarray}
 \Gamma_{XY}{}^{Z} & = & 
   {\textstyle\frac{1}{2}} g^{ZV} ( 2 \partial_{(X} g_{Y)V} - \partial_V g_{XY} ) \,, 
\\
 R_{XYZ}{}^{V} & = & 
 2 \partial_{[X} \Gamma_{Y]Z}{}^ V - 2 \Gamma_{[X|Z}{}^{W} \Gamma_{|Y]W}{}^{V} \,,
\\
 R_{XY} & = & R_{ZXY}{}^{Z} \,, 
\hspace{2em} 
 R = g^{XY} R_{XY} \,. 
\end{eqnarray}


\section{The Killing vectors of the target manifold}
\label{app:symmetrygroups}

In this appendix we show how the action of the symmetry groups $SO(5)$ or $SO(4,1)$, depending on whether the metric (\ref{generalmetric}) correspond to $S^4$ or $H^4$, is encoded in the general Killing vector (\ref{generalkilling}), whose parameters are subject to conditions (\ref{cond1}) - (\ref{cond3}). For the sake of shortheness we consider only the case of $a \neq 0$ and $b_X = 0$, which is the setting that leads directly to the stereographic representations of the metrics of $S^4$ and $H^4$.

The Eq.~(\ref{constraint}) can be used to solve for $c$, $c = \frac{\kappa}{12 a}$. From conditions (\ref{cond1}) - (\ref{cond3}) we may solve $\sigma$ and $\ell^X$,
\begin{equation}
 \sigma = 0 \,,
\hspace{2em}
 \ell^X = - \frac{ \kappa \lambda^X}{12 a^2}  \,.
\end{equation}
This leaves us with the ten parameters $\lambda^X$ and $\omega^{XY}$ as the free parameters of the Killing vector (\ref{generalkilling}), which takes the form
\begin{equation}
 k^X = \lambda^Y \left[
  \delta^{XY} \left( r^2 - \frac{\kappa}{12 a^2} \right)
 - 2 q^X q^Y  \right]
  + \omega^{XY} q^Y  \,.
\label{killinganonzero}
\end{equation}

If we make the immersions of $S^4$ and $H^4$ in $\mathbb{R}^5$, then they are defined by
\begin{equation}
 X^X X^X \pm (X^5)^2 = \pm 1
\end{equation}
where $X^{\hat{X}} = (X^X , X^5)$ are Cartesian coordinates in $\mathbb{R}^5$ and the upper (lower) signs hold for $S^4$ ($H^4$). The stereographic coordinates $q^X$ are defined by
\begin{equation}
 X^X = \frac{ 2 q^X}{1 \pm r^2} \,,
\hspace{2em}
X^5 = \frac{1 \mp r^2}{1 \pm r^2} \,,
\label{X}
\end{equation}
such that the inverse relations are
\begin{equation}
 q^X = \frac{X^X}{X^5 + 1} \,.
\end{equation}
The action of $SO(5)$ or $SO(4,1)$  in $\mathbb{R}^5$ can be represented by the infinitesimal coordinate transformation
\begin{equation}
 Y^{\hat{X}} = X^{\hat{X}} + t^{\hat{X}\hat{Y}} X^{\hat{Y}} \,,
\label{Y}
\end{equation}
where $t^{\hat{X}\hat{Y}}$ is a totally antisymmetric matrix for $SO(5)$ and a Lorentz generator for $SO(4,1)$,
\begin{equation}
 \mbox{$SO(5)$:} \;\; t^{\hat{X}\hat{Y}} = - t^{\hat{Y}\hat{X}} \,,
\hspace{2em}
 \mbox{$SO(4,1)$:} \;\; \left\{
 \begin{array}{l}
    t^{XY} = - t^{YX} \,, \\
    t^{5X} = + t^{X5} \,, \\
    t^{55} = 0 \,.
 \end{array}\right.
\end{equation}
This coordinate transformation is viewed on $S^4$ or $H^4$ as the new stereographic coordinate
\begin{equation}
 p^X = \frac{Y^X}{Y^5 + 1} \,.
\end{equation}
If we combine this equation with Eqs.~(\ref{X}) and (\ref{Y}) and expand in $t^{\hat{X}\hat{Y}}$ we arrive at the coordinate transformation on $S^4$ or $H^4$ given by
\begin{equation}
 p^X = q^X + t^{XY} q^Y 
   + {\textstyle\frac{1}{2}} t^{5Z} \left[ \delta^{ZX} ( r^2 \mp 1) - 2 q^X q^Z \right]\,.
\label{stereographictransformation}
\end{equation}

By comparing the coordinate transformation (\ref{stereographictransformation}) with the Killing vector (\ref{killinganonzero}) we may see that the Killing vector has the desired structure and it is only the sign of $\kappa$ what differentiates between $SO(5)$ and $SO(4,1)$.	


\end{document}